\documentclass[journal,twocolumn,10pt,twoside]{IEEEtranTCOM}
\newtheorem{definition}{{Definition}}
\newtheorem{theorem}{{Theorem}}

\newtheorem{corollary}{{Corollary}}
\newtheorem{remark}{{Remark}}

\usepackage{cite}
\usepackage{amsmath,amssymb,amsfonts}
\usepackage{algorithmic}
\usepackage{graphicx}
\usepackage{textcomp}
\usepackage{xcolor}

\normalsize

\ifCLASSINFOpdf
\else
\fi

\begin{document}

\title{On the Fundamental Limits of Integrated Sensing and Communications Under Logarithmic Loss}

\author{Jun Chen, Lei Yu, Yonglong Li, Wuxian Shi, Yiqun Ge, and Wen Tong}


\maketitle

\begin{abstract}
	We study a unified information-theoretic framework for integrated sensing and communications (ISAC), applicable to both monostatic and bistatic sensing scenarios. Special attention is given to the case where the sensing receiver (Rx) is required to produce a ``soft" estimate of the state sequence, with logarithmic loss serving as the performance metric. 	
	We derive lower and upper bounds on the capacity-distortion function, which delineates the fundamental tradeoff between communication rate and sensing distortion. These bounds coincide when the channel between the ISAC transmitter (Tx) and the communication Rx is degraded with respect to the channel between the ISAC Tx and the sensing Rx, or vice versa. 	
	Furthermore, we provide a complete characterization of the capacity-distortion function for an ISAC system that simultaneously transmits information over a binary-symmetric channel and senses  additive Bernoulli states through another binary-symmetric channel. The Gaussian counterpart of this problem is also explored, which, together with a state-splitting trick, fully determines the capacity-distortion-power function under the squared error distortion measure.
\end{abstract}	
	
	
	


%
\IEEEpeerreviewmaketitle

\section{Introduction}

Integrated sensing and communications (ISAC) \cite{FLiuetal22} represent a transformative approach that integrates sensing and communication functionalities within a unified system, enabling more efficient use of resources such as spectrum, power, and hardware. By leveraging the same infrastructure and signals for both tasks, ISAC not only reduces redundancy but also unlocks synergistic benefits that would otherwise be unattainable in separate systems. This integration is particularly significant in the context of emerging applications, such as autonomous vehicles, smart cities, and next-generation wireless networks, where simultaneous high-precision sensing and reliable communication are essential. ISAC promises to revolutionize these fields by enhancing performance, reducing costs, and paving the way for innovative applications that capitalize on its dual functionality.

\begin{figure}[htbp]
	\centerline{\includegraphics[width=5cm]{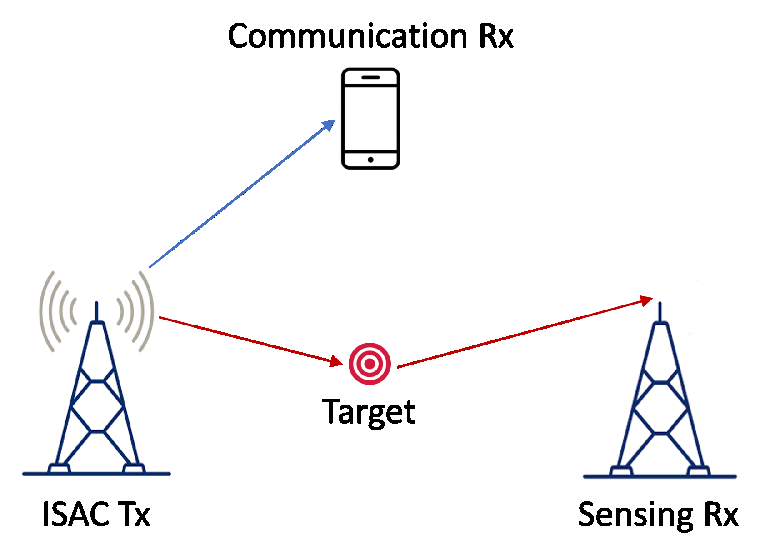}} \caption{Bistatic sensing (including monostatic sensing as a degenerate case).}
	\label{fig:system} 
\end{figure}

The development of ISAC systems presents a multitude of intriguing theoretical questions \cite{ALiuetal22,JW22,CWEB23,AWS23,NWSP24,WJ24}. At the heart of these questions is the need to understand the interplay between the communication and sensing functionalities, which is reflected in the fundamental tradeoff between communication rate and sensing distortion. This tradeoff encapsulates the challenge of optimizing the use of shared resources so that both high-quality sensing and efficient communication can be achieved. To address this, it is essential to bring together insights from two key theoretical areas: information theory, which provides the framework for understanding the performance limits of communication systems, and estimation theory, which focuses on how to best extract information about the sensed environment. The resulting comprehensive approach can lead to new methods of designing ISAC systems that effectively balance communication and sensing demands, thus enabling a wide range of practical applications.

The literature on ISAC primarily focuses on the monostatic sensing scenario, where the sensing receiver (Rx) is colocated with the ISAC transmitter (Tx) and, consequently, has knowledge of the channel input \cite{LXWHC23,XLCYHC23,AKWC24,JC24}. The key finding in this scenario is that the tradeoff between communication rate and sensing distortion is governed solely by constraints on the marginal distribution of the channel input. Operationally, this marginal distribution corresponds to a signaling strategy, which defines the constellation and the relative frequency with which each constellation point is used. It is important to note that the signaling strategy is only loosely connected to the communication rate. For instance, every non-degenerate linear code over $\mathrm{GF}(2)$ induces uniform binary signaling, meaning the actual code rate cannot be inferred from the signaling strategy alone. More generally, each signaling strategy has a maximum communication rate it can support; below this threshold, the specific rate can still be freely chosen. 
Therefore, in the monostatic sensing scenario, when the signaling strategy is fixed, communication rate and sensing distortion are effectively decoupled.

In the bistatic sensing scenario (see Fig. \ref{fig:system}), where the channel input is unknown to the sensing receiver (Rx), a natural question arises: does the decoupling principle observed in the monostatic setting still apply? As demonstrated in \cite{JGWWYC23},  this is generally not the case when the channel to the communication Rx is not degraded with respect to the channel to the sensing Rx. Intuitively, if the sensing Rx cannot decode the transmitted signal, a higher information content in the signal introduces greater uncertainty into the sensing task, leading to increased sensing distortion. In contrast, when the channel to the communication Rx is degraded with respect to the channel to the sensing Rx, the sensing Rx can decode the information-carrying signal intended for the communication Rx. As a result, the problem effectively reduces to the monostatic sensing scenario. In fact, monostatic sensing can be regarded as a special case of bistatic sensing where channel degradation holds trivially. This also explains why the decoupling phenomenon observed in monostatic sensing also manifests in the point-to-point setting where the Rx is tasked with both channel decoding and state estimation \cite{ZVM11}.

In this work, we follow \cite{JGWWYC23} by modeling the sensing target as a sequence of independent and identically distributed (i.i.d.) state variables. We extend the problem formulation in \cite{JGWWYC23} by relaxing the assumption that the state variables are directly observable by the communication Rx since this essentially excludes the important scenario where the sensing Rx is better informed than the communication Rx. 
On the other hand, our formulation is less general in one aspect compared to \cite{JGWWYC23}, as no common message is introduced for decoding by both the communication and sensing receivers. Nonetheless, the analysis reveals that even without enforcing a common message, it is often beneficial for the sensing Rx to partially decode the transmitted message and leverage it to improve state estimation, particularly when the channel to the sensing Rx is degraded relative to the channel to the communication Rx. Thus, this functionality is better left as a natural outcome rather than being explicitly imposed.

Beyond these relatively minor differences in system models, our work distinguishes itself from \cite{JGWWYC23} and the broader ISAC literature (with the exception of \cite{JC24}, which addresses the monostatic sensing scenario) by placing special emphasis  on the case where the sensing Rx is required to produce a posterior-distribution-like ``soft" estimate of the state sequence.  This focus is motivated by the fact that, in the Bayesian estimation framework, the posterior state distribution given the observation serves as a universal sufficient statistic, from which ``hard" reconstructions can be readily derived. 
To evaluate the quality of such a ``soft" estimate, we adopt logarithmic loss as the performance metric. Besides its practical movitations, the focus on state distribution estimation under logarithmic loss offers a mathematical advantage by faciliating the application of various extremal inequalities, leading to more explicit and conclusive results compared to other sensing tasks.

Below is a summary of our main contributions.
\begin{enumerate}
	\item We derive lower and upper bounds on the capacity-distortion function that delineates the fundamental information-theoretic limits of ISAC under logarithmic loss, along with a set of matching conditions. It is shown 
 that when the channel to the communication Rx is degraded with respect to the channel to the sensing Rx, the interplay between communication rate and sensing distortion occurs solely through the signaling strategy. This result extends the decoupling principle previously established for the monostatic sensing scenario \cite{JC24}.
	In contrast, when the channel to the sensing Rx is degraded with respect to the channel to the communication Rx, the decoupling principle no longer holds, reaffirming the observation made in \cite{JGWWYC23}.

	
	
	\item  We fully determine the capacity-distortion function for an ISAC system designed to simultaneously transmit information over a binary-symmetric channel and sense additive Bernoulli states through another binary-symmetric channel, regardless of the channel degradation order. The Gaussian counterpart of this problem is also explored, resulting in a complete characterization of the capacity-distortion-power function under the squared error distortion measure.
\end{enumerate}

The rest of this paper is organized as follows. Section \ref{sec:definition} provides the definition of the capacity-distortion function  along with a discussion of the connection between logarithm loss  and conventional distortion measures. Lower and upper bounds on the capacity-distortion function are presented in Section \ref{sec:bounds}, which coincide when the channel between the ISAC Tx and the communication Rx is degraded with respect to the channel between the ISAC Tx and the sensing Rx, or vice versa.
Conclusive results are derived in Section \ref{sec:specialcases} for both the binary and Gaussian cases. Section \ref{sec:conclusion} offers some concluding remarks.

Throughout this paper, we adopt standard notation for information measures: $H(\cdot)$ for entropy, $h(\cdot)$ for differential entropy, $I(\cdot;\cdot)$ for mutual information, and $H_b(\cdot)$ for the binary entropy function. Let $p_X$ denote the distribution of a random variable/vector $X$; specifically, $p_X$ is a probability mass function if $X$ is discret and  a probability density function if $X$ is contiunous. 
We use $\mathcal{B}(\alpha)$ and $\mathcal{N}(\mu,\sigma^2)$ to represent the Bernoulli distribution with parameter $\alpha$ and the Gaussian distribution with mean $\mu$ and variance $\sigma^2$, respectively. 
Let $\mathbb{E}[\cdot]$ denote the expectation 
operator, and let 
$\mathrm{var}(X|Y)$ represent the minimum mean sequared error in estimating $X$ given $Y$, i.e., $\mathrm{var}(X|Y):=\mathbb{E}[(X-\mathbb{E}[X|Y])^2]$. The probability of event $A$ is expressed as  $\mathbb{P}\{A\}$.
We define $a\oplus b$  as the modulo-2 addition and  $a*b$ as the binary convolution of $a$ and $b$ (i.e., $a*b:=(1-a)b+a(1-b)$).
Unless otherwise stated, the logarithm functions $\log(\cdot)$ and $\ln(\cdot)$ are assumed to have base $2$ and base $e$, respectively.

\section{Problem Definition}\label{sec:definition} 

Consider an ISAC system consisting of a Tx, a communication Rx, and a sensing Rx. The ISAC Tx encodes a message $M$, uniformly distributed over $\mathcal{M}$, 
into a codeword $X^n$ from codebook $\mathcal{C}$ and transmits $X^n$ through a broadcast channel\footnote{More precisely, the broadcast channel is specified by $p_{Y_1Y_2S|X}$, which can be factorized as $p_{Y_1Y_2|XS}p_S$ (i.e., the channel state is independent of the channel input).} 
$p_{Y_1Y_2|XS}$ with state distribution $p_S$. The communication Rx decodes the channel output 
$Y^n_1$
to produce a reconstructed message $\hat{M}$, while the sensing Rx uses the channel output 
$Y^n_2$ to generate an estimate $\hat{S}^n\in\hat{\mathcal{S}}^n$ of the state sequence $S^n$. Unless specified otherwise (see Section \ref{sec:Gaussian}), we assume that the input alphabet $\mathcal{X}$, the output alphabets $\mathcal{Y}_i$, $i=1,2$, and the state alphabet $\mathcal{S}$ all have finite cardinalities.

Let $d^{(n)}:\mathcal{S}^n\times\hat{\mathcal{S}}^{(n)}\rightarrow[-\infty,\infty]$, $n=1,2,\ldots$, be a sequence of distortion measures.  

\begin{definition}\label{def:capacity-distortion}
We say that a rate $R$ is achievable with respect to a distortion level $D$ 
if, for any $\epsilon>0$ and all sufficienty large $n$, there exist an encoding function $f^{(n)}:\mathcal{M}\rightarrow\mathcal{C}$, a decoding function $g^{(n)}_1:\mathcal{Y}^n_1\rightarrow\mathcal{M}$, and an estimation function $g^{(n)}_2:\mathcal{Y}^n_2\rightarrow\hat{\mathcal{S}}^{(n)}$ such that
\begin{align}
	&\frac{1}{n}\log|\mathcal{M}|\geq R-\epsilon,\\
	&\mathbb{P}\{\hat{M}\neq M\}\leq\epsilon,\\
	&\frac{1}{n}\mathbb{E}[d^{(n)}(S^n,\hat{S}^{(n)})]\leq D+\epsilon.\label{eq:distortion}
\end{align}
The maximum of such achievable rates $R$ is denoted by $C(D)$, which is referred to as the capacity-distortion function.
\end{definition}
\begin{remark}\label{remark:marginal}
	It is easy to verify that $C(D)$ depends on $p_{Y_1Y_2S|X}$ only through $p_{Y_1|X}$ and $p_{Y_2S|X}$.
\end{remark}

Special attention will be paid to the case where $\hat{\mathcal{S}}^{(n)}=\mathcal{P}(\mathcal{S}^n)$, with $\mathcal{P}(\mathcal{S}^n)$ representing the set of probability distributions over $\mathcal{S}^n$. In this case, $\hat{S}^{(n)}$ can be interpreted as a ``soft" estimate of $S^n$ given $Y^n_2$. Accordingly, we will denote $\hat{S}^{(n)}$ by $q(\cdot|Y^n_2)$. Two types of logarithmic loss are considered in this paper: sequence-wise logarithmic loss ($d^{(n)}=d^{(n)}_{\mathrm{seq}}$) and symbol-wise
logarithmic loss ($d^{(n)}=d^{(n)}_{\mathrm{sym}}$).

The sequence-wise logarithmic loss distortion measure is defined as
\begin{align*}
	d^{(n)}_{\mathrm{seq}}(s^n,\hat{s}^{(n)}):=\log\left(\frac{1}{q(s^n|y^n_2)}\right).
\end{align*}
Note that $\mathbb{E}[d^{(n)}(S^n,\hat{S}^{(n)})]=0$ if and only if $q(\cdot|Y^n_2)$ assigns probability $1$ to $S^n$ almost surely; such a ``soft" estimate is deemed perfect.

It is instructive to view $q(\cdot|Y^n_2)$ as a proxy for the posterior distribution $p_{S^n|Y^n_2}(\cdot|Y^n_2)$, which is actually the best ``soft" estimate of $S^n$ given $Y^n$ under sequence-wise logarithmic loss. Indeed, 
\begin{align}
	&\frac{1}{n}\mathbb{E}[d^{(n)}_{\mathrm{seq}}(S^n,\hat{S}^{(n)})]\nonumber\\
	&=\frac{1}{n}\sum\limits_{s^n\in\mathcal{S}^n,y^n_2\in\mathcal{Y}^n_2}p_{S^nY^n_2}(s^n,y^n_2)\log\left(\frac{1}{q(s^n|y^n_2)}\right)\nonumber\\
	&=\frac{1}{n}\sum\limits_{y^n_2\in\mathcal{Y}^n_2}p_{Y^n_2}(y^n_2)\sum\limits_{s^n\in\mathcal{S}^n}p_{S^n|Y^n_2}(s^n|y^n_2)\log\left(\frac{1}{q(s^n|y^n_2)}\right)\nonumber\\
	&=\frac{1}{n}\sum\limits_{y^n_2\in\mathcal{Y}^n_2}p_{Y^n_2}(y^n_2)D\left(p_{S^n|Y^n_2}(\cdot|y^n_2)\|q(\cdot|y^n_2)\right)\nonumber\\
	&\quad+\frac{1}{n}H(S^n|Y^n_2)\nonumber\\
	&\geq\frac{1}{n}H(S^n|Y^n_2),\label{eq:ws}
\end{align} 
and this lower bound is attained if and only if $q(\cdot|Y^n_2)=p_{S^n|Y^n_2}(\cdot|Y^n_2)$ almost surely.

Within the Bayesian estimation framework, $p_{S^n|Y^n_2}(\cdot|Y^n_2)$ serves as a universal sufficient statistic for $S^n$ given $Y^n_2$ due to the fact that $Y^n_2\leftrightarrow p_{S^n|Y^n_2}(\cdot|Y^n_2)\leftrightarrow S^n$ forms a Markov chain. The best ``hard" reconstruction of $S^n$ based on $Y^n_2$ 
in terms of a conventional sequence-wise distortion measure $\tilde{d}^{(n)}:\mathcal{S}^n\times\mathcal{S}^n\rightarrow[0,\infty]$ can be directly inferred from $p_{S^n|Y^n_2}(\cdot|Y^n_2)$ as follows:
\begin{align}
	\arg\min\limits_{\tilde{s}^n\in\mathcal{S}^n}\sum\limits_{s^n\in\mathcal{S}^n}p_{S^n|Y^n_2}(s^n|Y^n_2)\tilde{d}^{(n)}(s^n,\tilde{s}^n).\label{eq:softtohard}
\end{align}
Substituting $p_{S^n|Y^n_2}(\cdot|Y^n_2)$ with $q(\cdot|Y^n_2)$ in \eqref{eq:softtohard} yields a natural ``hard" reconstruction rule based on the ``soft" estimate $q(\cdot|Y^n_2)$. 

For symbol-wise logarithmic loss, the ``soft" estimate $S^{(n)}$ is restricted to the form $\prod_{t=1}^nq_t(\cdot|Y^n_2)$ with $q_t(\cdot|Y^n_2)\in\mathcal{P}(\mathcal{S})$, $t=1,2,\ldots,n$, and we define
\begin{align}
	d^{(n)}_{\mathrm{sym}}(s^n,\hat{s}^{(n)})&:=\log\left(\frac{1}{\prod_{t=1}^nq_t(s(t)|y^n_2)}\right)\nonumber\\
	&=\sum\limits_{t=1}^n\log\left(\frac{1}{q_t(s(t)|y^n_2)}\right).
\end{align}
Here, $q_t(\cdot|Y^n_2)$ can be interpreted as a proxy of the posterior distribution $p_{S(t)|Y^n_2}(\cdot|Y^n_2)$, $t=1,2,\ldots,n$. Similarly to \eqref{eq:ws}, we have
\begin{align}
	&\frac{1}{n}\mathbb{E}[d^{(n)}_{\mathrm{sym}}(S^n,\hat{S}^{(n)})]\nonumber\\
	&=\frac{1}{n}\sum\limits_{s^n\in\mathcal{S}^n,y^n_2\in\mathcal{Y}^n_2}p_{S^nY^n_2}(s^n,y^n_2)\log\left(\frac{1}{\prod_{t=1}^nq_t(s(t)|y^n_2)}\right)\nonumber\\
	&=\frac{1}{n}\sum\limits_{t=1}^n\sum\limits_{y^n_2\in\mathcal{Y}^n_2}p_{Y^n_2}(y^n_2)\sum\limits_{s^n\in\mathcal{S}^n}p_{S^n|Y^n_2}(s^n|y^n_2)\log\left(\frac{1}{q_t(s(t)|y^n_2)}\right)\nonumber\\
	&=\frac{1}{n}\sum\limits_{t=1}^n\sum\limits_{y^n_2\in\mathcal{Y}^n_2}p_{Y^n_2}(y^n_2)\sum\limits_{s\in\mathcal{S}}p_{S(t)|Y^n_2}(s|y^n_2)\log\left(\frac{1}{q_t(s(t)|y^n_2)}\right)\nonumber\\
	&=\frac{1}{n}\sum\limits_{t=1}^n\sum\limits_{y^n_2\in\mathcal{Y}^n_2}p_{Y^n_2}(y^n_2)\sum\limits_{s\in\mathcal{S}}p_{S(t)|Y^n_2}(s|y^n_2)\log\left(\frac{p_{S(t)|Y^n_2}(s|y^n_2)}{q_t(s(t)|y^n_2)}\right)\nonumber\\
	&\quad-\frac{1}{n}\sum\limits_{t=1}^n\sum\limits_{y^n_2\in\mathcal{Y}^n_2}p_{Y^n_2}(y^n_2)\sum\limits_{s\in\mathcal{S}}p_{S(t)|Y^n_2}(s|y^n_2)\log\left(p_{S(t)|Y^n_2}(s|y^n_2)\right)\nonumber\\
	&=\frac{1}{n}\sum\limits_{t=1}^n\sum\limits_{y^n_2\in\mathcal{Y}^n_2}p_{Y^n_2}(y^n_2)D\left(p_{S(t)|Y^n_2}(\cdot|y^n_2)\|q_t(\cdot|y^n_2)\right)\nonumber\\
	&\quad+\frac{1}{n}\sum\limits_{t=1}^nH(S(t)|Y^n_2)\nonumber\\
	&\geq\frac{1}{n}\sum\limits_{t=1}^nH(S(t)|Y^n_2),\label{eq:ss}
\end{align} 
and this lower bound is attained if and only if 
$q_t(\cdot|Y^n_2)=p_{S(t)|Y^n_2}(\cdot|Y^n_2)$, $t=1,2,\ldots,n$, almost surely.

Substituting $p_{S^n|Y^n_2}(\cdot|Y^n_2)$ with $\prod_{t=1}^nq_t(\cdot|Y^n_2)$ in \eqref{eq:softtohard} yields 
\begin{align}
	\arg\min\limits_{\tilde{s}^n\in\mathcal{S}^n}\sum\limits_{s^n\in\mathcal{S}^n}\left(\prod_{t=1}^nq_{t}(s(t)|Y^n_2)\right)\tilde{d}^{(n)}(s^n,\tilde{s}^n),\label{eq:softtohard2}
\end{align}
which reduces to the following symbol-wise ``hard" reconstruction rule
\begin{align}
	\arg\min\limits_{\tilde{s}\in\mathcal{S}}\sum\limits_{s\in\mathcal{S}}q_t(s|Y^n_2)\tilde{d}(s,\tilde{s}),\quad t=1,2,\ldots,n,
\end{align}
if $\tilde{d}^{(n)}(s^n,\tilde{s}^n)=\sum_{t=1}^n\tilde{d}(s(t),\tilde{s}(t))$ for some distortion measure $\tilde{d}:\mathcal{S}\times\mathcal{S}\in[0,\infty]$. Note that for any conventional symbol-wise distortion measure $\tilde{d}$, 
the ``hard" reconstruction based on $\prod_{t=1}^nq_t(\cdot|Y^n)$  coincides with that based on $q(\cdot|Y^n)$  if $q_t(\cdot|Y^n)$, $t=1,2,\ldots,n$, can be obtained from $q(\cdot|Y^n)$
 via marginalization.




Since a ``soft" estimate of the form $\prod_{t=1}^nq_t(\cdot|Y^n_2)$ remains valid for sequence-wise logarithmic loss, it follows that converse results established under sequence-wise logarithmic loss automatically hold under symbol-wise logarithmic loss, while achievability results established under symbol-wise logarithmic loss also apply to sequence-wise logarithmic loss.

Every achievable distortion level  under a conventional symbol-wise distortion measure $\tilde{d}$ implies an achievable distortion level under symbol-wise logarithmic loss, and consequently, also under sequence-wise logarithmic loss. Specifically, let $\tilde{S}^n$ denote a ``hard" reconstruction of $S^n$ based on $Y^n_2$ such that
\begin{align}
	\frac{1}{n}\sum\limits_{t=1}^n\mathbb{E}[\tilde{d}(S(t),\tilde{S}(t))]\leq\tilde{D};
\end{align}
by setting $q_t(\cdot|Y^n_2):=p_{S(t)|Y^n_2}(\cdot|Y^n_2)$, $t=1,2,\ldots$, we have
\begin{align}
	\frac{1}{n}\mathbb{E}[d^{(n)}_{\mathrm{sym}}(S^n,\hat{S}^{(n)})]&=\frac{1}{n}\sum\limits_{t=1}^nH(S(t)|Y^n_2)\nonumber\\
	&=\frac{1}{n}\sum\limits_{t=1}^n\left(H(S(t))-I(S(t);Y^n_2)\right)\nonumber\\
	&\leq\frac{1}{n}\sum\limits_{t=1}^n\left(H(S(t))-I(S(t);\tilde{S}(t))\right)\nonumber\\
	&\leq H(S)-R_S(\tilde{D}),
\end{align}
where $R_S(\tilde{D})$ is the rate-distortion function for the source distribution $p_S$ under the distortion measure $\tilde{d}$, i.e., 
\begin{align}
	R_S(\tilde{D})&:=\min\limits_{p_{\tilde{S}|S}}I(S;\tilde{S})\\
	&\mbox{s.t.}\quad \mathbb{E}[\tilde{d}(S,\tilde{S})]\leq \tilde{D}.
\end{align}
In other words, if a distortion level $\tilde{D}$ is achievable under the distortion measure $\tilde{d}$ by a given scheme, then the product of the posterior distributions $p_{S(t)|Y^n_2}(\cdot|Y^n_2)$, $t=1,2,\ldots,n$, induced by that scheme achieves a distortion level no greater than $H(S)-R_S(\tilde{D})$ under both symbol-wise logarithmic loss and sequence-wise logarithmic loss.








Logarithmic loss is widely used as penalty function in the theory of learning and prediction \cite[Chapter 9]{CBL06}. Its prominence in information theory can be attributed to the seminal work of Courtade and Weissman \cite{CW14}. Our preceding discussion
demonstrates that, in the ISAC setting, logarithmic loss serves as a natural performance metric for ``soft" sensing, as it accomodates a wide range of 
Bayesian estimation tasks.

\section{Bounds and Matching Conditions}\label{sec:bounds}

In this section, we establish lower and upper bounds on the capacity-distortion function under logarithmic loss. Additionally, it is shown that these bounds coincide when a degradation order exists between the channel to the communication Rx and the channel to the sensing Rx.

\subsection{Lower Bound}

\begin{theorem}\label{thm:lowerbound}
	Under both sequence-wise and symbol-wise logarithm loss, 
	\begin{align}
	C(D)\geq\underline{C}(D),
	\end{align}
where
\begin{align}
	\underline{C}(D):=&\max\limits_{p_{UX}}\min\{I(X;Y_1),I(X;Y_1|U)+I(U;Y_2)\}\label{eq:achievability}\\
	&\mbox{s.t.}\quad H(S|U,Y_2)\leq D,
\end{align}
with the joint distribution $p_{UXY_1Y_2S}$ factorized as $p_{UX}p_{Y_1Y_2S|X}$.
\end{theorem}
\begin{remark}
	 For the auxiliary random variable $U$ in the definition of $\underline{C}(D)$, the cardinality bound on its alphabet $\mathcal{U}$ can be established as follows. 	 
	 By the support lemma \cite[p. 310, Lemma 3.4]{CK81}, there is no loss of generality in assuming  $|\mathcal{U}|\leq|\mathcal{X}|+1$ as it suffices to preserve $p_X$, $H(Y_1|U)+H(Y_2|U)$, and $H(S|U,Y_2)$. This also justifies the use of ``$\max$" in \eqref{eq:achievability}.
\end{remark}
\begin{IEEEproof}
The achievability of the lower bound $\underline{C}(D)$ is based on the standard superposition coding scheme \cite{EGK11}. It can be viewed as an adaption of \cite[Theorem 1]{JGWWYC23} to the logarithmic loss setting.
Thus, only a sketch of the proof is provided here.

It suffices to focus on symbol-wise logarithmic loss, as the corresponding achievabiity result also applies to sequence-wise logarithmic loss. 
Generate $2^{nR_1}$ codewords $\{U^n(m_1)\}_{m_1=1}^{2^{nR_1}}$ using $p_U$, and for each $U^n(m_1)$, generate $2^{nR_2}$ codewords $\{X^n(m_1,m_2)\}_{m_2=1}^{2^{nR_2}}$ using $p_{X|U}$. Given a message $M:=(M_1,M_2)$ uniformly distributed over $\{1,2,\ldots,2^{nR_1}\}\times\{1,2,\ldots,2^{nR_2}\}$, the ISAC Tx sends $X^n(M_1,M_2)$ through the channel. We choose 
\begin{align}
	&R_1\approx\min\{I(U;Y_1),I(U;Y_2)\},\\
	&R_2\approx I(X;Y_1|U),
\end{align}
such that, with high probability, the sensing Rx can decode $M_1$, while the communication Rx can decode $(M_1,M_2)$. The sensing Rx then sets $q_t(\cdot|Y^n_2):=p_{S|UY_2}(\cdot|U(M_1,t),Y_2(t))$, where $u(M_1,t)$ denotes the $t$-entry of $U^n(M_1)$, $t=1,2,\ldots,n$. This resulting ``soft" estimate $\hat{S}^{(n)}:=\prod_{t=1}^nq_t(\cdot|Y^n_2)$ satisfies
\begin{align}
	\frac{1}{n}\mathbb{E}[d^{(n)}_{\mathrm{sym}}(S^n,\hat{S}^{(n)})]\approx H(S|U,Y_2)
\end{align}
with high probability.
\end{IEEEproof}

\subsection{Upper Bound}

\begin{theorem}\label{thm:upperbound}
Under symbol-wise logarithmic loss,
	\begin{align}
		C(D)\leq\overline{C}_{\mathrm{sym}}(D),\label{eq:ssupper}
	\end{align} 
and under sequence-wise logarithmic loss,
\begin{align}
	C(D)\leq\overline{C}_{\mathrm{seq}}(D),\label{eq:wsupper}
\end{align} 
where
\begin{align}
	\overline{C}_{\mathrm{sym}}(D):=&\max\limits_{p_{UVX}}\min\{I(X;Y_1),I(X;Y_1|U)+I(U;Y_2),\nonumber\\
	&\hspace{0.4in}I(X;Y_1,Y_2|U,V)+I(U,V;Y_2)\}\label{eq:ssconverse}\\
	&\mbox{s.t.}\quad H(S|U,V,Y_2)\leq D,
\end{align}
and
\begin{align}
\overline{C}_{\mathrm{seq}}(D):=&\max\limits_{p_{UVX}}\min\{I(X;Y_1),I(X;Y_1|U)+I(U;Y_2),\nonumber\\
&\hspace{0.35in}I(X;Y_1,Y_2,S|U,V)+I(U,V;Y_2)\}\label{eq:wsconverse}\\
&\mbox{s.t.}\quad H(S|U,V,Y_2)\leq D,
\end{align}
with the joint distribution $p_{UVXSY_1Y_2}$ factorized as $p_{UVX}p_{Y_1Y_2S|X}$.
\end{theorem}
\begin{remark}
For the auxiliary random variables $U$ and $V$ in the definition of $\overline{C}_{\mathrm{sym}}$, the cardinality bounds on their alphabets $\mathcal{U}$ and $\mathcal{V}$ can be established as follows.
By the support lemma \cite[p. 310, Lemma 3.4]{CK81}, there is no loss of generality in assuming  $|\mathcal{U}|\leq|\mathcal{X}|+2$ as it suffices to preserve $p_X$, $H(Y_1|U)+H(Y_2|U)$, $H(Y_1,Y_2|U,V)+H(Y_2|U,V)$, and $H(S|U,V,Y_2)$. Moreover, no loss of generality is incured by assuming $|\mathcal{V}|\leq |\mathcal{U}||\mathcal{X}|+1$ as it suffices to preserve $p_{UX}$,  $H(Y_1,Y_2|U,V)+H(Y_2|U,V)$, and $H(S|U,V,Y_2)$. The same cardinality bounds apply to $U$ and $V$ in the definition of $\overline{C}_{\mathrm{seq}}(D)$.
This also justifies the use of ``$\max$" in \eqref{eq:ssconverse} and \eqref{eq:wsconverse}.
\end{remark}
\begin{IEEEproof}
	See Appendix \ref{app:upperbound}.
\end{IEEEproof}



It can be shown via a timesharing argument that $C(D)$ is concave in $D$. Similarly, $\underline{C}(D)$, $\overline{C}_{\mathrm{sym}}(D)$, and $\overline{C}_{\mathrm{seq}}(D)$ are also concave in $D$, as the timesharing variable can be obsorbed into the auxiliary random vaiable $U$.


\subsection{Matching Conditions}

The standard notions of channel degradedness are adopted in this paper. Specifically, $p_{Y_1|X}$ is said to be physically degraded with respect to $p_{Y_2|X}$ if $p_{Y_1Y_2|X}=p_{Y_2|X}p_{Y_1|Y_2}$, and is said to be stochastically degraded with respect to $p_{Y_2|X}$ if there exists a transition matrix $\{q(y_1|y_2)\}_{y_1\in\mathcal{Y}_1,y_2\in\mathcal{Y}_2}$  such that $p_{Y|X}(y|x)=\sum_{y'\in\mathcal{Y}'}p_{Y'|X}(y'|x)q(y|y')$ for all $x\in\mathcal{X}$ and $y\in\mathcal{Y}$. The case where $p_{Y_2|X}$ is degraded with respect to $p_{Y_1|X}$ can be defined in a similar manner.

It is evident that physical degradedness implies stochastic degradedness, but the converse does not hold in general. However, we will demonstrate that these two notions of channel degradedness are, in fact, equivalent when it comes to characterizing 
the capacity-distortion function. This mirrors the equivalence observed in the conventional broadcast channel setting. A subtle yet crucial aspect here is ensuring that the conversion from stochastic degradedness to physical degradedness does not affect the sensing task.

Let $p_{Y_1|X}$, $p_{Y_2|X}$, and $p_{Y_2S|X}$ be the conditional distributions induced by the given broadcast channel $p_{Y_1Y_2S|X}$. Depending on whether $p_{Y_1|X}$ is stochastically degraded with respect to $p_{Y_2|X}$ or the other way around, we can construct $p'_{Y_1Y_2|X}$ by coupling $p_{Y_1|X}$ and $p_{Y_2|X}$ such that $X\leftrightarrow Y_2\leftrightarrow Y_1$ or $X\leftrightarrow Y_1\leftrightarrow Y_2$ forms a Markov chain under $p'_{Y_1Y_2|X}$. Furthermore, we can construct $p'_{Y_1Y_2S|X}$ by coupling $p'_{Y_1Y_2|X}$ and $p_{Y_2S|X}$ such that $Y_1\leftrightarrow(X,Y_2)\leftrightarrow S$ forms a Markov chain under $p'_{Y_1Y_2S|X}$. 
The constructed broadcast channel $p'_{Y_1Y_2S|X}$ possesses the desired physical degradedness structure. Moreover, in view of Remark \ref{remark:marginal}, it has the same capacity-distortion function as  the original broadcast channel $p_{Y_1Y_2S|X}$ because 
$p'_{Y_1|X}=p_{Y_1|X}$ and $p'_{Y_2S|X}=p_{Y_2S|X}$. Henceforth, we will no longer distinguish between the two notions of channel degradedness and will freely apply the properties of physical degradedness even when the given broadcast channel only satisfies stochastic degradedness.




\begin{corollary}\label{cor:degraded1}
If $p_{Y_1|X}$ is degraded with resepct to $p_{Y_2|X}$, then under both sequence-wise and symbol-wise logarithmic loss,
\begin{align}
	C(D)=&\max\limits_{p_{X}}I(X;Y_1)\\
	&\mbox{s.t.}\quad H(S|X,Y_2)\leq D.
\end{align}
\end{corollary}
\begin{IEEEproof}
Choosing $U=X$ shows
\begin{align}
	\underline{C}(D)\geq&\max\limits_{p_{X}}\min\{I(X;Y_1),I(X;Y_2)\}\\
&\mbox{s.t.}\quad H(S|X,Y_2)\leq D.
\end{align}
Since $p_{Y_1|X}$ is degraded with resepct to $p_{Y_2|X}$, it follows that $I(X;Y_1)\leq I(X;Y_2)$. Consequently, we have
\begin{align}
	\underline{C}(D)\geq&\max\limits_{p_{X}}I(X;Y_1)\\
	&\mbox{s.t.}\quad H(S|X,Y_2)\leq D.
\end{align}
Moreover, $H(S|U,V,Y_2)\geq H(S|X,Y_2)$ for $p_{UVXY_1Y_2S}$ factorized as $p_{UVX}p_{Y_1Y_2S|X}$. Therefore,
\begin{align}
	\overline{C}_{\mathrm{sym}}(D)\leq&\max\limits_{p_{X}}I(X;Y_1)\\
	&\mbox{s.t.}\quad H(S|X,Y_2)\leq D,
\end{align}
and
\begin{align}
	\overline{C}_{\mathrm{seq}}(D)\leq&\max\limits_{p_{X}}I(X;Y_1)\\
	&\mbox{s.t.}\quad H(S|X,Y_2)\leq D.
\end{align}
Invoking Theorems \ref{thm:lowerbound} and \ref{thm:upperbound} proves Corollary \ref{cor:degraded1}.
	\end{IEEEproof}

Corollary \ref{cor:degraded1} indicates that when $p_{Y_1|X}$ is degraded with respect to 
$p_{Y_2|X}$, the tradeoff between communication rate and sensing distortion is governed by the adjustment of the input distribution $p_X$. Practically, this is realized through the choice of signaling strategy or waveform design. For a fixed 
$p_X$, communication rate and sensing distortion are decoupled in the sense that any communication rate above  $I(X;Y_1)$ and any sensing distortion below  $H(S|X,Y_1)$
can be independently selected. 
Intuitively, since the sensing Rx can decode the message intended for the communication Rx due to the favorable channel degradation order, it is the statistical properties of the input, rather than the exact communication rate, that determine the sensing performance.
This result generalizes the decoupling principle previously established for monostatic sensing \cite[Theorem 1]{JC24}, which corresponds to the degenerate scenario where $X$ is a function of $Y_2$, making the degradedness of $p_{Y_1|X}$ relative to $p_{Y_2|X}$ trivially satisfied.




\begin{corollary}\label{cor:degraded2}
	If $p_{Y_2|X}$ is degraded with resepct to $p_{Y_1|X}$, then under symbol-wise logarithmic loss,
\begin{align}
	C(D)=&\max\limits_{p_{UX}}I(X;Y_1|U)+I(U;Y_2)\\
	&\mbox{s.t.}\quad H(S|U,Y_2)\leq D,
\end{align}
with the joint distribution $p_{UXY_1Y_2S}$ factorized as $p_{UX}p_{Y_1Y_2S|X}$. Furthermore, if $p_{Y_1Y_2S|X}=p_{Y_1|X}p_{Y_2|Y_1S}p_S$, then the same conclusion holds  under sequence-wise logarithmic loss,
\end{corollary}
\begin{remark}
	Without loss of generality, we can assume that  the alphabet $\mathcal{U}$ of the auxiliary random variable $U$ in the statement of Corollary \ref{cor:degraded2} satisfies the cardinality bound $|\mathcal{U}|\leq|\mathcal{X}|+1$.
\end{remark}
\begin{remark}
	The condition $p_{Y_1Y_2S|X}=p_{Y_1|X}p_{Y_2|Y_1S}p_S$ implies that $X\leftrightarrow Y_1\leftrightarrow (Y_2,S)$ forms a Markov chain, which is a stronger condition than the physical degradedness $X\leftrightarrow Y_1\leftrightarrow Y_2$.
\end{remark}
\begin{IEEEproof}
Since $p_{Y_2|X}$ is degraded with respect to $p_{Y_1|X}$, it follows that
	\begin{align}
		I(X;Y_1|U)+I(U;Y_2)&\leq I(X;Y_1|U)+I(U;Y_1)\nonumber\\&=I(U,X;Y_1)\nonumber\\
		&=I(X;Y_1)
	\end{align} 
	if $U\leftrightarrow X\leftrightarrow (Y_1,Y_2)$ forms a Markov chain. As a consequence, 
	\begin{align}
		\underline{C}(D)=&\max\limits_{p_{UX}}I(X;Y_1|U)+I(U;Y_2)\\
		&\mbox{s.t.}\quad H(S|U,Y_2)\leq D,
	\end{align}
	with the joint distribution $p_{UXY_1Y_2S}$ factorized as $p_{UX}p_{Y_1Y_2S|X}$.
	The degradedness of $p_{Y_2|X}$ relative to $p_{Y_1|X}$ also implies that
\begin{align}
	I(X;Y_1,Y_2|U,V)=I(X;Y_1|U,V)
\end{align}
if $(U,V)\leftrightarrow X\leftrightarrow(Y_1,Y_2)$ forms a Markov chain. Therefore, we have
\begin{align}
	\overline{C}_{\mathrm{sym}}(D)\leq&\max\limits_{p_{UVX}} I(X;Y_1|U,V)+I(U,V;Y_2)\\
	&\mbox{s.t.}\quad H(S|U,V,Y_2)\leq D,
\end{align}
with the joint distribution $p_{UVXY_1Y_2S}$ factorized as $p_{UVX}p_{Y_1Y_2S|X}$. Consolidating $U$ and $V$ into a single auxiliary random variable reveals that
\begin{align}
	\overline{C}_{\mathrm{sym}}(D)\leq\underline{C}(D).
\end{align}
This directly leads to the first statement of  Corollary \ref{cor:degraded2}, in light of Theorems \ref{thm:lowerbound} and \ref{thm:upperbound}.

Now we proceed to prove the second statement of Corollary \ref{cor:degraded2}. The condition
$p_{Y_1Y_2S|X}=p_{Y_1|X}p_{Y_2|Y_1S}p_S$ implies that
\begin{align}
	I(X;Y_1,Y_2,S|U,V)=I(X;Y_1|U,V)
\end{align}
if $(U,V)\leftrightarrow X\leftrightarrow(Y_1,Y_2,S)$ forms a Markov chain. Therefore, we have
\begin{align}
	\overline{C}_{\mathrm{seq}}(D)\leq&\max\limits_{p_{UVX}} I(X;Y_1|U,V)+I(U,V;Y_2)\\
	&\mbox{s.t.}\quad H(S|U,V,Y_2)\leq D,
\end{align}
with the joint distribution $p_{UVXY_1Y_2S}$ factorized as $p_{UVX}p_{Y_1Y_2S|X}$. The remainder of the proof follows the same reasoning as for the first statement.
\end{IEEEproof}
	




Corollary \ref{cor:degraded2} suggests that when $p_{Y_2|X}$ is degraded with respect to 
$p_{Y_1|X}$, the relationship between communication rate and sensing distortion is influenced by more than just the input distribution. Even with $p_X$ fixed, different choices of $U$ can directly impact the achievable communication rate and sensing distortion. This is because $I(U;Y_2)$ represents the amount of information that can be decoded by both the communication Rx and the sensing Rx, while $I(X;Y_1|U)$ corresponds to the information that is only decodable by the communication Rx. The more information decoded by the sensing Rx, the easier the sensing task tends to become. However, this comes at the cost of reducing the information that can be sent to the communication Rx, as additional redundancy is required to ensure the transmitted information  can withstand the less favorable channel conditions and remain  decodable at the sensing Rx.




\section{Special Cases}\label{sec:specialcases}

Unlike conventional distortion metrics, logarithmic loss enables the single-letter characterizations or bounds of the capacity-distortion function to be expressed purely in terms of information measures, which facilitates the application of extremal inequalities for their evaluation. We illustrate this with two special channel models and further highlight the significance of logarithmic loss by demonstrating how it yields conclusive results for conventional distortion metrics.


\subsection{Binary Case}

Consider the case where the channel from the ISAC Tx to the communication Rx is given by $Y_1=X\oplus Z_1$ and the channel to the sensing Rx is given by $Y_2=X\oplus Z_2\oplus S$. It is assumed that $Z_1\sim\mathcal{B}(\beta_1)$, $Z_2\sim\mathcal{B}(\beta_2)$, and $S\sim\mathcal{B}(\beta_S)$ with $\beta_1,\beta_2,\beta_S\in(0,\frac{1}{2})$; moreover,  $Z_1$, $Z_2$, and $S$ are independent of $X$ while $Z_2$ and $S$ are mutually independent. The capacity-distortion function for this channel model is denoted by $C_{\mathrm{B}}(D)$.


If $\beta_1\in[\beta_2*\beta_S,\frac{1}{2}]$, then $p_{Y_1|X}$ is degraded with respect to $p_{Y_2|X}$; so it follows by Corollary \ref{cor:degraded1} that under symbol-wise logarithmic loss, 
\begin{align}
	C_{\mathrm{B}}(D)=&\max\limits_{p_{X}}I(X;Y_1)\label{eq:binary1}\\
	&\mbox{s.t.}\quad H(S|X,Y_2)\leq D.\label{eq:binary2}
\end{align}
If $\beta_1\in[0,\beta_2*\beta_S]$, then $p_{Y_2|X}$ is degraded with respect to $p_{Y_1|X}$; so it follows by Corollary \ref{cor:degraded2} that under symbol-wise logarithmic loss, 
\begin{align}
	C_{\mathrm{B}}(D)=&\max\limits_{p_{UX}}I(X;Y_1|U)+I(U;Y_2)\label{eq:binary3}\\
	&\mbox{s.t.}\quad H(S|U,Y_2)\leq D,\label{eq:binary4}
\end{align}
with the joint distribution $p_{UXY_1Y_2S}$ factorized as $p_{UX}p_{Y_1Y_2S|X}$. 
As $U\leftrightarrow X\leftrightarrow (Y_1,Y_2,S)$ forms a Markov chain, we have
\begin{align}
	H(S|Z_2\oplus S)=H(S|X,Y_2)\leq H(S|U,Y_2)\leq H(S).
\end{align}
The distortion constraint becomes unsatisfiable when  $D<H(S|Z_2\oplus S)$ and inactive when $D>H(S)$. 
Note that 
\begin{align}
	H(S|Z_2\oplus S)&=H(S,Z_2\oplus S)-H(Z_2\oplus S)\nonumber\\
	&=H(S)+H(Z_2)-H(Z_2\oplus S)\nonumber\\
	&=H_b(\beta_2)+H_b(\beta_S)-H_b(\beta_2*\beta_S)
\end{align}
and $H(S)=H_b(\beta_S)$.
Therefore, it suffices to consider $D\in[H_b(\beta_2)+H_b(\beta_S)-H_b(\beta_2*\beta_S),H_b(\beta_S)]$.

\begin{figure}[htbp]
	\centerline{\includegraphics[width=6cm]{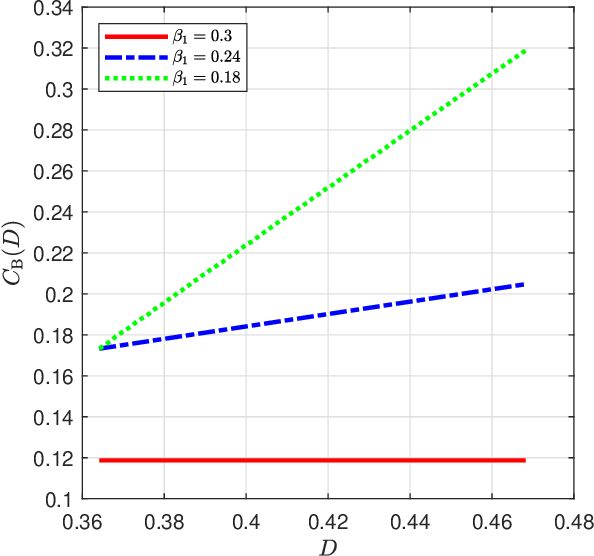}} \caption{Plots of $C_{\mathrm{B}}(D)$ under sybmol-wise logarithmic loss with $\beta_2=0.2$ and $\beta_S=0.1$ are shown for $\beta_1=0.3$, $\beta_1=0.24$, and $\beta_1=0.18$, corresponding to cases 1), 2) and 3) in Theorem \ref{thm:binary}, respectively. The plots for $\beta_1=0.3$ and $\beta_1=0.18$ also apply to sequence-wise logarithmic loss, while the plot for $\beta_1=0.24$ serves as a lower bound on $C_{\mathrm{B}}(D)$ under sequence-wise logarithmic loss.}
	\label{fig:binary} 
\end{figure}

	Even with the single-letter characterization in \eqref{eq:binary1}--\eqref{eq:binary4}, the exact evaluation of $C_{\mathrm{B}}(D)$ remains highly non-trivial. We address this challenge by establishing an extremal inequality for certain linear combinations of binary entropy functions, which leads to the result below (see also Fig. \ref{fig:binary}).

\begin{theorem}\label{thm:binary}
Under symbol-wise logarithmic loss, $C_{\mathrm{B}}(D)$ admits the following explicit characterization:
	
	\begin{enumerate}
		\item If $\beta_1\in[\beta_2*\beta_S,\frac{1}{2})$, then		
	\begin{align}
		C_{\mathrm{B}}(D)=1-H_b(\beta_1)\label{eq:binarycase1}
	\end{align}
for $D\in[H_b(\beta_2)+H_b(\beta_S)-H_b(\beta_2*\beta_S),H_b(\beta_S)]$.
		
		\item If $\beta_1\in(\beta_2,\beta_2*\beta_S)$, 	then
		\begin{align}
			C_{\mathrm{B}}(D)&=\frac{H_b(\beta_2*\beta_S)-H_b(\beta_1)}{H_b(\beta_2*\beta_S)-H_b(\beta_2)}\left(D-H_b(\beta_S)\right)\nonumber\\
			&\quad+1-H_b(\beta_1)	\label{eq:binarycase2}
		\end{align}
		for $D\in[H_b(\beta_2)+H_b(\beta_S)-H_b(\beta_2*\beta_S),H_b(\beta_S)]$. 
		
		\item If $\beta_1\in(0,\beta_2]$, then
		\begin{align}
			C_{\mathrm{B}}(D)&=1-H_b(\beta_1)+H_b(\alpha_D*\beta_1)\nonumber\\
			&\quad-H_b(\alpha_D*\beta_2*\beta_S)\label{eq:binarycase3}
		\end{align}
		for $D\in[H_b(\beta_2)+H_b(\beta_S)-H_b(\beta_2*\beta_S),H_b(\beta_S)]$, where $\alpha_D$ is the unique number in $[0,\frac{1}{2}]$ satisfying 
		\begin{align}
			H_b(\beta_S)+H_b(\alpha_D*\beta_2)-H_b(\alpha_D*\beta_2*\beta_S)=D.
		\end{align}	
	\end{enumerate}
	Moreover, cases 1) and 3) also hold under  sequence-wise logarithmic loss.
\end{theorem}
\begin{remark}
A computable characterization of $C_{\mathrm{B}}(D)$ under sequence-wise logarithmic loss is unknown in case 2); nevertheless, 
since sequence-wise logarithmic loss  is more relaxed than symbol-wise logarithmic loss, we must have
\begin{align}
	C_{\mathrm{B}}(D)&\geq\frac{H_b(\beta_2*\beta_S)-H_b(\beta_1)}{H_b(\beta_2*\beta_S)-H_b(\beta_2)}\left(D-H_b(\beta_S)\right)\nonumber\\
	&\quad+1-H_b(\beta_1)	
\end{align}
for $D\in[H_b(\beta_2)+H_b(\beta_S)-H_b(\beta_2*\beta_S),H_b(\beta_S)]$. 
\end{remark}
\begin{IEEEproof}
For case 1), the single-letter characterization of $C_{\mathrm{B}}(D)$ under symbol-wise logarithmic loss is given by \eqref{eq:binary1} and \eqref{eq:binary2}. 
Regardless of the choice of $p_X$, we have $H(S|X,Y_2)=H_b(\beta_2)+H_b(\beta_S)-H_b(\beta_2*\beta_S)$, and consequently, the constraint in \eqref{eq:binary2} is trivially satisifed for $D\in[H_b(\beta_2)+H_b(\beta_S)-H_b(\beta_2*\beta_S),H_b(\beta_S)]$. Moreover, the objective function in \eqref{eq:binary1} attains its maximum value of $1-H_b(\beta_1)$ when $X\sim\mathcal{B}(\frac{1}{2})$. This proves \eqref{eq:binarycase1}.

For cases 2) and 3), the single-letter characterization of $C_{\mathrm{B}}(D)$ under symbol-wise logarithmic loss is given by \eqref{eq:binary3} and \eqref{eq:binary4}. Let $X=U\oplus\Delta$, where $U\sim\mathcal{B}(\frac{1}{2})$ and $\Delta\sim\mathcal{B}(\alpha)$ are mutually independent and also independent of $(Z_1,Z_2,S)$. It can be verified that 
\begin{align}
	&I(X;Y_1|U)+I(U;Y_2)\nonumber\\
	&=H(Y_1|U)-H(Y_1|U,X)+H(Y_2)-H(Y_2|U)\nonumber\\
	&=H(\Delta\oplus Z_1)-H(Z_1)+H(Y_2)-H(\Delta\oplus Z_2\oplus S)\nonumber\\
	&=1-H_b(\beta_1)+H_b(\alpha*\beta_1)-H_b(\alpha*\beta_2*\beta_S)
\end{align}
and
\begin{align}
	&H(S|U,Y_2)\nonumber\\
	&=H(S|\Delta\oplus Z_2\oplus S)\nonumber\\
	&=H(S,\Delta\oplus Z_2\oplus S)-H(\Delta\oplus Z_2\oplus S)\nonumber\\
	&=H(S)+H(\Delta\oplus Z_2)-H(\Delta\oplus Z_2\oplus S)\nonumber\\
	&=H_b(\beta_S)+H_b(\alpha*\beta_2)-H_b(\alpha*\beta_2*\beta_S).
\end{align}
As $\alpha$ varies from $0$ to $\frac{1}{2}$, $H(S|U,Y_2)$ increases from $H_b(\beta_2)+H_b(\beta_S)-H_b(\beta_2*\beta_S)$ to $H_b(\beta_S)$.
Choosing $\alpha=\alpha_D$ ensures that $H(S|U,Y_2)=D$, and consequently, the distortion constraint in \eqref{eq:binary4} is satisfied. Therefore, the corresponding $I(X;Y_1|U)+I(U;Y_2)$ provides a lower bound on $C_{\mathrm{B}}(D)$, i.e.,  $C_{\mathrm{B}}(D)\geq R_{\mathrm{B}}(D)$, where
\begin{align}
	R_{\mathrm{B}}(D):=1-H_b(\beta_1)+H_b(\alpha_D*\beta_1)-H_b(\alpha_D*\beta_2*\beta_S).
\end{align}
Since $C_{\mathrm{B}}(D)$ is concave in $D$, it follows that the upper concave envelope of $R_{\mathrm{B}}(D)$, denoted by $\overline{R}_{\mathrm{B}}(D)$, is also a lower bound on 
$C_{\mathrm{B}}(D)$. We establish an extremal inequality in Appendix \ref{app:binary}, which implies $C_{\mathrm{B}}(D)=\overline{R}_{\mathrm{B}}(D)$. 
The problem now reduces to computing $\overline{R}_{\mathrm{B}}(D)$. It turns out that in case 3), $R_{\mathrm{B}}(D)$ is concave in $D$, and thus coincides with $\overline{R}_{\mathrm{B}}(D)$. This proves \eqref{eq:binarycase3}. However, in case 2), $R_{\mathrm{B}}(D)$ is convex in $D$, and consequently, $\overline{R}_{\mathrm{B}}(D)$ is the line segment connecting the two endpoints of $R_{\mathrm{B}}(D)$ 
at $D=H_b(\beta_2)+H_b(\beta_S)-H_b(\beta_2*\beta_S)$ and $D=H_b(\beta_S)$, respectively. The operational implication is  that in case 2), $C_{\mathrm{B}}(D)$ is achieved by timesharing between the two coding schemes associated with $U=X$ and $U=0$. This proves \eqref{eq:binarycase2}. 	
Further details are provided in Appendix \ref{app:binary}.
	
It remains to address the statement regarding sequence-wise logarithmic loss. 
In case 1), since $p_{Y_1|X}$ is degraded with respect to $p_{Y_2|X}$,  both types of logarithmic loss yield the same $C_{\mathrm{B}}(D)$ according to Corollary \ref{cor:degraded1}. In case 3), we can express $Z_2$ as
$Z_2=Z_1\oplus W$ with $W\sim\mathcal{B}(\frac{\beta_2-\beta_1}{1-2\beta_1})$ and assume that $X$, $Z_1$, $W$, and $S$ are mutually independent. This construction ensures   $p_{Y_1Y_2S|X}=p_{Y_1|X}p_{Y_2|Y_1S}p_S$. Consequently, by Corollary \ref{cor:degraded2},  $C_{\mathrm{B}}(D)$ under sequence-wise logarithmic loss coincides with that under symbol-wise logarithmic loss. 
This completes the proof of Theorem \ref{thm:binary}.
	\end{IEEEproof}

\subsection{Gaussian Case}\label{sec:Gaussian}

Consider the case where the channel from the ISAC Tx to the communication Rx is given by $Y_1=X+ Z_1$ and the channel to the sensing Rx is given by $Y_2=X+ Z_2+ S$. It is assumed that $Z_1\sim\mathcal{N}(0,N_1)$, $Z_2\sim\mathcal{N}(0,N_2)$, and $S\sim\mathcal{N}(0,N_S)$ with $N_1,N_2,N_S>0$; moreover,  $Z_1$, $Z_2$, and $S$ are independent of $X$ while $Z_2$ and $S$ are mutually independent. We modify Definition \ref{def:capacity-distortion} by incorporating an additional average power contraint
\begin{align}
	\frac{1}{n}\sum\limits_{t=1}^n\mathbb{E}[X^2(t)]\leq P+\epsilon\label{eq:power}
\end{align}
and denote the resulting  capacity-distortion-power function  by $C_{\mathrm{G}}(D,P)$.

It is worth mentioning that logarithmic loss can be naturally extended from discrete to continuous random variables by replacing probability mass functions with probability density functions. Consequently, differential entropy takes the place of entropy in the relevant expressions. In particular, \eqref{eq:ws} and \eqref{eq:ss} become
\begin{align}
\frac{1}{n}\mathbb{E}[d^{(n)}_{\mathrm{seq}}(S^n,\hat{S}^{(n)})]\geq\frac{1}{n}h(S^n|Y^n_2)
\end{align}
and
\begin{align}
	\frac{1}{n}\mathbb{E}[d^{(n)}_{\mathrm{sym}}(S^n,\hat{S}^{(n)})]\geq\frac{1}{n}\sum\limits_{t=1}^nh(S(t)|Y^n_2),
\end{align}
respectively. Note that logarithmic loss may not always be non-negative for continuous random variables.

If $N_1\in[N_2+N_S,\infty)$, then $p_{Y_1|X}$ is degraded with respect to $p_{Y_2|X}$; a simple extension\footnote{Both Corollaries \ref{cor:degraded1} and \ref{cor:degraded2} can be extended to the Gaussian case. In particular, the achievability proof follows from  the standard discretization procedure, while the converse can be established using the weak convergence argument in \cite[Appendix II]{GN14}.} of Corollary \ref{cor:degraded1} shows that under symbol-wise logarithmic loss, 
\begin{align}
	C_{\mathrm{G}}(D,P)=&\max\limits_{p_{X}}I(X;Y_1)\label{eq:Gaussian1}\\
	&\mbox{s.t.}\quad h(S|X,Y_2)\leq D,\label{eq:Gaussian2}\\
	&\hspace{0.3in}\mathbb{E}[X^2]\leq P.\label{eq:Gaussian3}
\end{align}
If $N_1\in(0,N_2+N_S]$, then $p_{Y_2|X}$ is degraded with respect to $p_{Y_1|X}$; a simple extension of Corollary \ref{cor:degraded2} shows that under symbol-wise logarithmic loss, 
\begin{align}
	C_{\mathrm{G}}(D,P)=&\max\limits_{p_{UX}}I(X;Y_1|U)+I(U;Y_2)\label{eq:Gaussian4}\\
	&\mbox{s.t.}\quad h(S|U,Y_2)\leq D,\label{eq:Gaussian5}\\
	&\hspace{0.3in} \mathbb{E}[X^2]\leq P,\label{eq:Gaussian6}
\end{align}
with the joint distribution $p_{UXY_1Y_2S}$ factorized as $p_{UX}p_{Y_1Y_2S|X}$.
As $U\leftrightarrow X\leftrightarrow (Y_1,Y_2,S)$ forms a Markov chain, we have
\begin{align}
	h(S|Z_2+S)=h(S|X,Y_2)\leq h(S|U,Y_2)\leq h(S|Y_2).
\end{align}
The distortion constraint becomes unsatisfiable when  $D<h(S|Z_2+S)$ and inactive when $D>\max_{p_X:\mathbb{E}[X^2]\leq P}h(S|Y_2)$. 
Note that 
\begin{align}
	h(S|Z_2+S)&=\frac{1}{2}\log\left(2\pi e\mathrm{var}(S|Z_2+S)\right)\nonumber\\
	&=\frac{1}{2}\log\left(\frac{2\pi eN_2N_S}{N_2+N_S}\right)
\end{align}
and 
\begin{align}
	\max\limits_{p_X:\mathbb{E}[X^2]\leq P}h(S|Y_2)&\leq\max\limits_{p_X:\mathbb{E}[X^2]\leq P}\frac{1}{2}\log(2\pi e\mathrm{var}(S|Y_2))\nonumber\\
	&\leq\frac{1}{2}\log\left(\frac{2\pi e(P+N_2)N_S}{P+N_2+N_S}\right).
\end{align}
Therefore, it suffices to consider $D\in[\frac{1}{2}\log(\frac{2\pi eN_2N_S}{N_2+N_S}),\frac{1}{2}\log(\frac{2\pi e(P+N_2)N_S}{P+N_2+N_S})]$.

\begin{figure}[htbp]
	\centerline{\includegraphics[width=6cm]{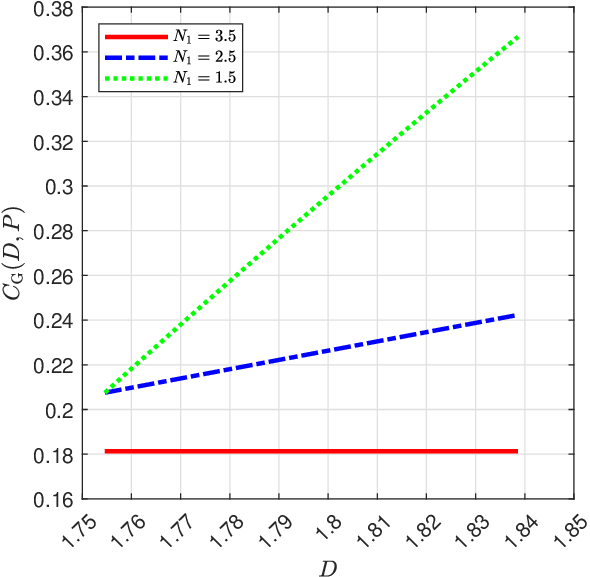}} \caption{Plots of $C_{\mathrm{G}}(D,P)$ with $P=1$, $N_2=2$, and $N_S=1$ are shown for $N_1=3.5$, $N_1=2.5$, ad $N_1=1.5$, corresponding to cases 1), 2) and 3) in Theorem \ref{thm:Gaussian1}, respectively. They apply to both sequence-wise  and symbol-wise logarithmic loss. The plots for $N_1=3.5$ and $N_1=1.5$ depict the exact values of $C_{\mathrm{G}}(D,P)$, while the plot for $N_1=2.5$ serves as a lower bound on $C_{\mathrm{G}}(D,P)$.}
	\label{fig:Gaussian1} 
\end{figure}

We establish a variant of the entropy power inequality and leverage it to perform an exact evaluation of the single-letter characterization of $C_{\mathrm{G}}(D,P)$ in \eqref{eq:Gaussian1}--\eqref{eq:Gaussian6}, which leads to the result below (see also Fig. \ref{fig:Gaussian1}).

\begin{theorem}\label{thm:Gaussian1}
	Under both sequence-wise and symbol-wise logarithmic loss, $C_{\mathrm{G}}(D,P)$ admits the following explicit characterization or bound:
	\begin{enumerate}
		\item If $N_1\in[N_2+N_S,\infty)$, then
		\begin{align}
			C_{\mathrm{G}}(D,P)=\frac{1}{2}\log\left(\frac{P+N_1}{N_1}\right)\label{eq:Gaussiancase1}
		\end{align}
	for $D\in[\frac{1}{2}\log(\frac{2\pi eN_2N_S}{N_2+N_S}),\frac{1}{2}\log(\frac{2\pi e(P+N_2)N_S}{P+N_2+N_S})]$.
		\item If $N_1\in(N_2,N_2+N_S)$, then
		\begin{align}
			C_{\mathrm{G}}(D,P)&\geq\frac{1}{2}\log\left(\frac{P+N_1}{N_1}\right)+\frac{\log\left(\frac{(P+N_1)(N_2+N_S)}{N_1(P+N_2+N_S)}\right)}{\log\left(\frac{(P+N_2)(N_2+N_S)}{(P+N_2+N_S)N_2}\right)}\nonumber\\
			&\quad\times\left(D-\frac{1}{2}\log\left(\frac{2\pi e(P+N_2)N_S}{P+N_2+N_S}\right)\right)\label{eq:Gaussiancase2}
		\end{align}
		for $D\in[\frac{1}{2}\log(\frac{2\pi eN_2N_S}{N_2+N_S}),\frac{1}{2}\log(\frac{2\pi e(P+N_2)N_S}{P+N_2+N_S})]$.
		\item If $N_1\in(0,N_2]$, then
			\begin{align}
			C_{\mathrm{G}}(D,P)&=\frac{1}{2}\log\Big(\frac{P+N_2+N_S}{2\pi eN_1N^2_S}(2\pi e(N_1-N_2)N_S\nonumber\\
			&\hspace{0.6in}+(N_2+N_S-N_1)2^{2D})\Big)\label{eq:Gaussiancase3}
		\end{align}
	for $D\in[\frac{1}{2}\log(\frac{2\pi eN_2N_S}{N_2+N_S}),\frac{1}{2}\log(\frac{2\pi e(P+N_2)N_S}{P+N_2+N_S})]$.
	\end{enumerate}
\end{theorem}
\begin{remark}
	The lower bound \eqref{eq:Gaussiancase2} in case 2) is not jointly concave in $(D,P)$. As a result, it can be further improved by performing simultaneous timesharing with respect to $D$ and $P$. Interestingly, even with this enhancement, the lower bound remains suboptimal. Indeed, if multiple power levels  are involved, then the timesharing variable, denoted by $Q$, is not independent of $X$ and, consequently, not independent of $Y_2$. This leads to the inequality
	\begin{align}
		&I(X;Y_1|U,Q)+I(U;Y_2|Q)\nonumber\\
		&<I(X;Y_1|U,Q)+I(Q,U;Y_2).
	\end{align}
Thus, merging $U$ and $Q$ into a single auxiliary random variable can strictly outperform pure timesharing.
\end{remark}
\begin{IEEEproof}
	For case 1), the single-letter characterization of $C_{\mathrm{G}}(D,P)$ under symbol-wise logarithmic loss is given by \eqref{eq:Gaussian1}--\eqref{eq:Gaussian3}. Regardless of the choice of $p_X$, we have $h(S|X,Y_2)=\frac{1}{2}\log(\frac{2\pi eN_2N_S}{N_2+N_S})$, and consequently, the constraint in \eqref{eq:Gaussian2} is trivially satisifed for $D\in[\frac{1}{2}\log(\frac{2\pi eN_2N_S}{N_2+N_S}),\frac{1}{2}\log(\frac{2\pi e(P+N_2)N_S}{P+N_2+N_S})]$.	
	 Moreover, the objective function in \eqref{eq:Gaussian1} attains its maximum value of $\frac{1}{2}\log(\frac{P+N_1}{N_1})$ subject to the power constraint in \eqref{eq:Gaussian3} when $X\sim\mathcal{N}(0,P)$. This proves \eqref{eq:Gaussiancase1}.
	
	For cases 2) and 3), the single-letter characterization of $C_{\mathrm{G}}(D,P)$ under symbol-wise logarithmic loss is given by \eqref{eq:Gaussian4}--\eqref{eq:Gaussian6}. Let $X=U+\Delta$, where $U\sim\mathcal{N}(0,P-P')$ and $\Delta\sim\mathcal{N}(0,P')$ are mutually independent and also independent of $(Z_1,Z_2,S)$. We have $X\sim\mathcal{N}(0,P)$, which satisfies
	the power constraint in \eqref{eq:Gaussian6}.
	Moreover, it can be verified that 
	\begin{align}
		&I(X;Y_1|U)+I(U;Y_2)\nonumber\\
		&=h(Y_1|U)-h(Y_1|U,X)+h(Y_2)-h(Y_2|U)\nonumber\\
		&=h(\Delta+ Z_1)-h(Z_1)+h(Y_2)-h(\Delta+ Z_2+ S)\nonumber\\
		&=\frac{1}{2}\log\left(\frac{(P'+N_1)(P+N_2+N_S)}{N_1(P'+N_2+N_S)}\right)
	\end{align}
	and
	\begin{align}
		h(S|U,Y_2)&=h(S|\Delta+Z_2+ S)\nonumber\\
		&=\frac{1}{2}\log(2\pi e\mathrm{var}(S|\Delta+Z_2+ S))\nonumber\\
		&=\frac{1}{2}\log\left(\frac{2\pi e(P'+N_2)N_S}{P'+N_2+N_S}\right).
	\end{align}
	As $P'$ varies from $0$ to $P$, $h(S|U,Y_2)$ increases from $\frac{1}{2}\log(\frac{2\pi eN_2N_S}{N_2+N_S})$ to $\frac{1}{2}\log(\frac{2\pi e(P+N_2)N_S}{P+N_2+N_S})$.
	Choosing 
	\begin{align}
		P'=\frac{N_S}{2\pi eN_S2^{-2D}-1}-N_2
	\end{align}
	ensures that $h(S|U,Y_2)=D$, and consequently, the distortion constraint in \eqref{eq:Gaussian5} is satisfied. Therefore, the corresponding $I(X;Y_1|U)+I(U;Y_2)$ provides a lower bound on $C_{\mathrm{G}}(D,P)$, i.e.,  $C_{\mathrm{G}}(D,P)\geq R_{\mathrm{G}}(D,P)$, where
	\begin{align}
	R_{\mathrm{G}}(D,P)&:=\frac{1}{2}\log\Big(\frac{P+N_2+N_S}{2\pi eN_1N^2_S}(2\pi e(N_1-N_2)N_S\nonumber\\
	&\hspace{0.9in}+(N_2+N_S-N_1)2^{2D})\Big).
	\end{align}
	Since $C_{\mathrm{G}}(D,P)$ is concave in $(D,P)$, it follows that the upper concave envelope of $R_{\mathrm{G}}(D,P)$, denoted by $\overline{R}_{\mathrm{G}}(D,P)$, is also a lower bound on 
	$C_{\mathrm{G}}(D,P)$. 	
	In Appendix \ref{app:Gaussian}, we establish a variant of the entropy power inequality, which implies $C_{\mathrm{G}}(D,P)=\overline{R}_{\mathrm{G}}(D,P)=R_{\mathrm{G}}(D,P)$ in case 3). 
	This proves \eqref{eq:Gaussiancase3}. It can be verified that
	\begin{align}
		\frac{\partial R^2_{\mathrm{G}}(D,P)}{\partial D^2}=\frac{(4\pi e\ln 2)(N_1-N_2)(N_2+N_S-N_1)N_S2^{2D}}{(2\pi e(N_1-N_2)N_S+(N_2+N_S-N_1)2^{2D})^2}.
	\end{align}
Therefore, in case 2), for a fixed $P$, $R_{\mathrm{G}}(D,P)$ is convex in $D$,  and consequently, $\overline{R}_{\mathrm{G}}(D,P)$ is bounded from below by the line segment  connecting the two endpoints of $R_{\mathrm{G}}(D,P)$ 
	at $D=\frac{1}{2}\log(\frac{2\pi eN_2N_S}{N_2+N_S})$ and $D=\frac{1}{2}\log(\frac{2\pi e(P+N_2)N_S}{P+N_2+N_S})$, respectively. 
	This proves \eqref{eq:Gaussiancase2}.
	
	
	Now we proceed to consider sequence-wise logarithmic loss. In case 1), $p_{Y_1|X}$ is degraded with respect to $p_{Y_2|X}$. In case 3), we can express $Z_2$ as
	$Z_2=Z_1+ W$ with $W\sim\mathcal{N}(0,N_2-N_1)$ and assume that $X$, $Z_1$, $W$, and $S$ are mutually independent; this construction ensures   $p_{Y_1Y_2S|X}=p_{Y_1|X}p_{Y_2|Y_1S}p_S$. Therefore, adapting Corollaries \ref{cor:degraded1} and \ref{cor:degraded2} to the Gaussian channel model shows that in both cases, $C_{\mathrm{G}}(D,P)$ under sequence-wise logarithmic loss coincides with that under symbol-wise logarithmic loss. Moreover, in case 2), the lower bound on $C_{\mathrm{G}}(D,P)$ under symbol-wise logarithmic loss also applies under sequence-wise logarithmic loss, as the latter is a more relaxed distortion measure. This completes the proof of Theorem \ref{thm:Gaussian1}. 	
\end{IEEEproof}

As discussed in Section \ref{sec:definition}, there exist intimate connections between logarithmic loss and conventional distortion measures. Here, we take the widely used squared error distortion measure as an example and examine the corresponding capacity-distortion-power function $C'_{\mathrm{G}}(D,P)$ for the Gaussian channel model, defined by  substituting $d^{(n)}(S^n,\hat{S}^{(n)})$ with $\sum_{t=1}^n(S(t)-\hat{S}(t))^2$ in \eqref{eq:distortion} and incorprating the average power constraint \eqref{eq:power}.

The single-letter characterization of $C'_{\mathrm{G}}(D,P)$ under the squared error distortion measure closely resembles its counterpart under symbol-wise logarithmic loss. Specifically, for $N_1\in[N_2+N_S,\infty)$, 
\begin{align}
	C'_{\mathrm{G}}(D,P)=&\max\limits_{p_{X}}I(X;Y_1)\label{eq:mse1}\\
	&\mbox{s.t.}\quad \mathrm{var}(S|X,Y_2)\leq D,\label{eq:mse2}\\
	&\hspace{0.3in}\mathbb{E}[X^2]\leq P,\label{eq:mse3}
\end{align}
while for $N_1\in(0,N_2+N_S]$, 
\begin{align}
	C'_{\mathrm{G}}(D,P)=&\max\limits_{p_{UX}}I(X;Y_1|U)+I(U;Y_2)\label{eq:mse4}\\
	&\mbox{s.t.}\quad \mathrm{var}(S|U,Y_2)\leq D,\label{eq:mse5}\\
	&\hspace{0.3in} \mathbb{E}[X^2]\leq P,\label{eq:mse6}
\end{align}
with the joint distribution $p_{UXY_1Y_2S}$ factorized as $p_{UX}p_{Y_1Y_2S|X}$.
As $U\leftrightarrow X\leftrightarrow (Y_1,Y_2,S)$ forms a Markov chain, we have
\begin{align}
	\mathrm{var}(S|Z_2+S)=\mathrm{var}(S|X,Y_2)\leq \mathrm{var}(S|U,Y_2)\leq \mathrm{var}(S|Y_2).
\end{align}
The distortion constraint becomes unsatisfiable when  $D<\mathrm{var}(S|Z_2+S)$ and inactive when $D>\max_{p_X:\mathbb{E}[X^2]\leq P}\mathrm{var}(S|Y_2)$. 
Note that 
\begin{align}
	\mathrm{var}(S|Z_2+S)=\frac{N_2N_S}{N_2+N_S}
\end{align}
and 
\begin{align}
	\max\limits_{p_X:\mathbb{E}[X^2]\leq P}\mathrm{var}(S|Y_2)\leq\frac{(P+N_2)N_S}{P+N_2+N_S}.
\end{align}
Therefore, it suffices to consider $D\in[\frac{N_2N_S}{N_2+N_S},\frac{(P+N_2)N_S}{P+N_2+N_S}]$.

\begin{figure}[htbp]
	\centerline{\includegraphics[width=6cm]{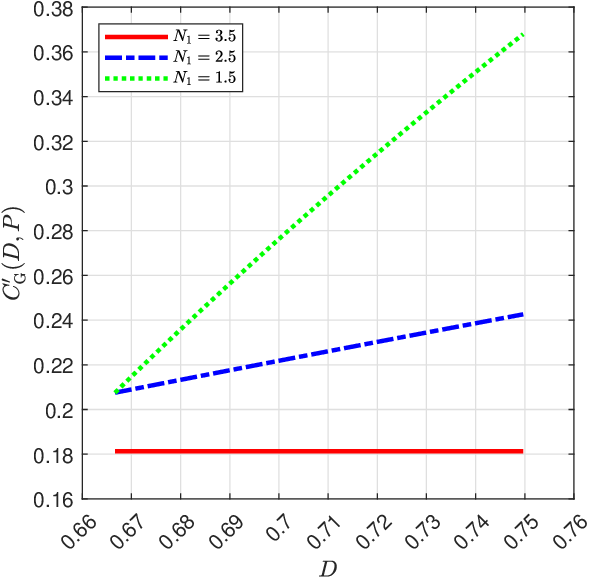}} \caption{Plots of $C'_{\mathrm{G}}(D,P)$ under the squared error distortion measure with $P=1$, $N_2=2$, and $N_S=1$ are shown for $N_1=3.5$, $N_1=2.5$, and $N_1=1.5$. Here, $N_1=3.5$ corresponds to case 1), while both $N_1=2.5$ and $N_1=1.5$ correspond to case 2) in Theorem \ref{thm:Gaussian2}. Different from its logarithmic loss counterpart, the plot for $N_1=2.5$ depicts the exact valus of $C'_{\mathrm{G}}(D,P)$, rather than just a lower bound.}
	\label{fig:Gaussian2} 
\end{figure}

Building upon Theorem \ref{thm:Gaussian1}, we derive a complete characterization of $C'_{\mathrm{G}}(D,P)$ under the squared error distortion measure (see also Fig. \ref{fig:Gaussian2}). This result may seem surprising, given that its counterpart under logarithmic loss remains unknown for $N_1\in(N_2,N_2+N_S)$. However, as we will see, the key insight is that under the squared error distortion measure, this challenging case can be reduced to the simpler case $N_1\in(0,N_2]$ via a state-splitting trick.

\begin{theorem}\label{thm:Gaussian2}
	Under the squared error distortion measure,
	$C'_{\mathrm{G}}(D,P)$ admits the following explicit characterization:
	\begin{enumerate}
		\item If $N_1\in[N_2+N_S,\infty)$, then
		\begin{align}
			C'_{\mathrm{G}}(D,P)=\frac{1}{2}\log\left(\frac{P+N_1}{N_1}\right)\label{eq:msecase1}
		\end{align}
	for $D\in[\frac{N_2N_S}{N_2+N_S},\frac{(P+N_2)N_S}{P+N_2+N_S}]$.
		\item If $N_1\in(0,N_S+N_2)$, then
	\begin{align}
		C'_{\mathrm{G}}(D,P)&=\frac{1}{2}\log\Big(\frac{P+N_2+N_S}{N_1N^2_S}((N_1-N_2)N_S\nonumber\\
		&\hspace{0.7in}+(N_2+N_S-N_1)D)\Big)\label{eq:msecase2}
	\end{align}
		for $D\in[\frac{N_2N_S}{N_2+N_S},\frac{(P+N_2)N_S}{P+N_2+N_S}]$.
	\end{enumerate}
\end{theorem}
\begin{IEEEproof}
For case 1), the single-letter characterization of $C'_{\mathrm{G}}(D,P)$ under the squared error distortion measure is given by \eqref{eq:mse1}--\eqref{eq:mse3}. Regardless of the choice of $p_X$, we have $\mathrm{var}(S|X,Y_2)=\frac{N_2N_S}{N_2+N_S}$, and consequently, the constraint in \eqref{eq:mse2} is trivially satisifed for $D\in[\frac{N_2N_S}{N_2+N_S},\frac{(P+N_2)N_S}{P+N_2+N_S}]$.	
Moreover, the objective function in \eqref{eq:mse1} attains its maximum value of $\frac{1}{2}\log(\frac{P+N_1}{N_1})$ subject to the power constraint in \eqref{eq:mse3} when $X\sim\mathcal{N}(0,P)$. This proves \eqref{eq:msecase1}.

For case 2), the single-letter characterization of $C'_{\mathrm{G}}(D,P)$ under the squared error distortion measure is given by \eqref{eq:mse4}--\eqref{eq:mse6}. Let $X=U+\Delta$, where $U\sim\mathcal{N}(0,P-P')$ and $\Delta\sim\mathcal{N}(0,P')$ are mutually independent and also independent of $(Z_1,Z_2,S)$. We have $X\sim\mathcal{N}(0,P)$, which satisfies
the power constraint in \eqref{eq:Gaussian6}.
Moreover, it can be verified that 
\begin{align}
	&I(X;Y_1|U)+I(U;Y_2)\nonumber\\
	&=\frac{1}{2}\log\left(\frac{(P'+N_1)(P+N_2+N_S)}{N_1(P'+N_2+N_S)}\right)
\end{align}
and
\begin{align}
	\mathrm{var}(S|U,Y_2)&=\mathrm{var}(S|\Delta+Z_2+ S)\nonumber\\
	&=\frac{(P'+N_2)N_S}{P'+N_2+N_S}.
\end{align}
As $P'$ varies from $0$ to $P$, $\mathrm{var}(S|U,Y_2)$ increases from $\frac{N_2N_S}{N_2+N_S}$ to $\frac{(P+N_2)N_S}{P+N_2+N_S}$.
Choosing 
\begin{align}
	P'=\frac{N_SD}{N_S-D}-N_2
\end{align}
ensures that $\mathrm{var}(S|U,Y_2)=D$, and consequently, the distortion constraint in \eqref{eq:mse5} is satisfied. Therefore, the corresponding $I(X;Y_1|U)+I(U;Y_2)$ provides a lower bound on $C'_{\mathrm{G}}(D,P)$, i.e.,  $C'_{\mathrm{G}}(D,P)\geq R'_{\mathrm{G}}(D,P)$, where
\begin{align}
R'_{\mathrm{G}}(D,P)&=\frac{1}{2}\log\Big(\frac{P+N_2+N_S}{N_1N^2_S}((N_1-N_2)N_S\nonumber\\
&\hspace{1.0in}+(N_2+N_S-N_1)D)\Big).
\end{align}

To establish the tightness of this lower bound, 
we first consider the subcase $N_1\in(0,N_2]$. Since $\mathrm{var}(S|U,Y_2)\leq D$ implies $h(S|U,Y_2)\leq\frac{1}{2}\log(2\pi e D)$, it follows that $C'_{\mathrm{G}}(D,P)$ under the squared error distortion measure is upper-bounded by $C_{\mathrm{G}}(\frac{1}{2}\log(2\pi e D),P)$ under symbol-wise logarithmic loss. In light of \eqref{eq:Gaussiancase3}, the latter coincides with $R'_{\mathrm{G}}(D,P)$. 
 This proves \eqref{eq:msecase2} for $N_1\in(0,N_2]$.

It remains to treat the subcase $N_1\in(N_2,N_2+N_S)$. We can split $S$ into two independent random variables, $S'$ and $S''$, where $S'\sim\mathcal{N}(0,N_2+N_S-N_1)$ and $S''\sim\mathcal{N}(0,N_1-N_2)$. Moreover,  $S'$ and $S''$ are assumed to be independent of $(X,Z_2)$. Let
\begin{align}
	Q(t)&:=S'(t)-\mathbb{E}[S'(t)|S(t)]\nonumber\\
	&=S'(t)-\frac{N_2+N_S-N_1}{N_S}S(t),\quad t=1,2,\ldots,n.
\end{align}
It can be verified that  $Q(t)\sim\mathcal{N}(0,\frac{(N_1-N_2)(N_2+N_S-N_1)}{N_S})$, $t=1,2,\ldots,n$, and $Q^n$ is independent of $(X^n,Y^n_2,S^n)$. Therefore,
\begin{align}
	\frac{1}{n}\sum\limits_{t=1}^n\mathrm{var}(S'(t)|Y^n_2)&=\frac{(N_2+N_S-N_1)^2}{nN^2_S}\sum\limits_{t=1}^n\mathrm{var}(S(t)|Y^n_2)\nonumber\\
	&\quad+\frac{(N_1-N_2)(N_2+N_S-N_1)}{N_S}.\label{eq:equivalence}
\end{align}
Now consider a new ISAC system with $S'$ as the target state and $Z_2+S''$ as the effective noise for the channel to the sensing Rx. In light of \eqref{eq:equivalence}, 
the original ISAC system with distortion constraint $D$ is equivalent to 
the new ISAC system with distortion constraint
\begin{align}
D':=\frac{(N_2+N_S-N_1)((N_1-N_2)(N_S-D)+N_SD)}{N^2_S}. 
\end{align}
Since the variance of the effective noise $Z_2+S''$ is $N_1$, the new ISAC system falls into the previously solved subcase, for which the capacity-distortion-power function has been characterized. 
Substituting $N_2$, $N_S$, and $D$ with $N_1$, $N_2+N_S-N_1$, and $D'$, respectively, in \eqref{eq:msecase2}
completes the proof for the subcase $N_1\in(N_2,N_2+N_S)$.
\end{IEEEproof}

\section{Conclusion}\label{sec:conclusion}

We have studied the fundamental limits of ISAC under logarithmic loss. Specifically, lower and upper bounds on the capacity-distortion function are established and are shown to coincide  when the channels to the communicaiton Rx and the sensing Rx exhibit a degradation order. Explicit results are derived for the binary case and the Gaussian case with the aid of various extremal inequalities.

A complete characterization of the capacity-distortion function in the absence of a channel degradation order remains an open problem of significant interest. Furthermore, evaluating the capacity-distortion function and its associated bounds appears to be highly non-trivial, highlighting the need for the development of new extremal inequalities and efficient numerical algorithms.


Two types of logarithmic loss are considered in this work: sequence-wise and symbol-wise.
However, the distinction between them requires further clarification. Thus far, they have led to the same capacity-distortion function in cases where definitive results have been obtained. It is yet unclear whether this alignment stems from the specific channel conditions considered or represents a more general phenomenon.





\appendices

\section{Proof of Theorem \ref{thm:upperbound}}\label{app:upperbound}

By Definition \ref{def:capacity-distortion}, for any $\epsilon>0$ and all sufficiently large $n$, there exist an encoding function $f^{(n)}:\mathcal{M}\rightarrow\mathcal{C}$, a decoding function $g^{(n)}_1:\mathcal{Y}^n_1\rightarrow\mathcal{M}$, and an estimation function $g^{(n)}_2:\mathcal{Y}^n_2\rightarrow\hat{\mathcal{S}}^{(n)}$ such that
\begin{align}
	&\frac{1}{n}\log|\mathcal{M}|\geq C(D)-\epsilon,\\
	&\mathbb{P}\{\hat{M}\neq M\}\leq\epsilon,\\
	&\frac{1}{n}\mathbb{E}[d^{(n)}(S^n,\hat{S}^{(n)})]\leq D+\epsilon.
\end{align}

First consider the case of symbol-wise logarithmic loss (i.e., $d^{(n)}=d^{(n)}_{\mathrm{sym}}$).
Let $T$ be uniformly distributed over $\{1,2,\ldots,n\}$ and independent of $(M,X^n,S^n,Y^n_1,Y^n_2)$, and set\footnote{This identification is justified by the fact that $p_{Y_1(T)Y_2(T)S(T)|X(T)}$ coincides with the given broadcast channel $p_{Y_1Y_2S|X}$.} $X:=X(T)$,  $Y_1:=Y_1(T)$, $Y_2:=Y_2(T)$, and $S:=S(T)$.
It can be shown by following the standard steps of the converse argument for the channel coding theorem (see, e.g., \cite[Chapter 3.1.4]{EGK11}) that
\begin{align}
	\frac{1}{n}\log|\mathcal{M}|\leq I(X;Y_1)+\delta_n(\epsilon),\label{eq:constraint1}
\end{align} 
where $\delta_n(\epsilon)$ tends to $0$ as  $n\rightarrow\infty$ and $\epsilon\rightarrow 0$.
Moreover, setting $U(t):=(Y^{t-1}_1,Y^n_{2,t+1})$, $t=1,2,\ldots,n$, and $U:=(U(T),T)$, we have
\begin{align}
	\frac{1}{n}\log|\mathcal{M}|	&\leq\frac{1}{n}I(M;Y^n_1)+\delta_n(\epsilon)\nonumber\\
	&=\frac{1}{n}\sum\limits_{t=1}^nI(M;Y_1(t)|Y^{t-1}_1)+\delta_n(\epsilon)\nonumber\\
	&\leq\frac{1}{n}\sum\limits_{t=1}^n(I(X(t),Y^n_{2,t+1};Y_1(t)|Y^{t-1}_1)\nonumber\\
	&\hspace{0.5in}+I(Y^n_{2,t+1};Y_2(t)))+\delta_n(\epsilon)\nonumber\\
	&=\frac{1}{n}\sum\limits_{t=1}^n(I(X(t);Y_1(t)|U(t))+I(U(t);Y_2(t)))\nonumber\\
	&\quad+\delta_n(\epsilon)\nonumber\\
	&\leq I(X;Y_1|U)+I(U;Y_2)+\delta_n(\epsilon).\label{eq:Y_1}
\end{align}
Substituting $Y_1(t)$ with $(Y_1(t),Y_2(t))$, $t=1,2,\ldots,n$, in the derivation of \eqref{eq:Y_1} yields 
\begin{align}
	\frac{1}{n}\log|\mathcal{M}|	
	&\leq\frac{1}{n}\sum\limits_{t=1}^n(I(X(t);Y_1(t),Y_2(t)|U(t),V(t))\nonumber\\
	&\hspace{0.5in}+I(U(t),V(t);Y_2(t)))+\delta_n(\epsilon)\nonumber\\
	&\leq I(X;Y_1,Y_2|U,V)+I(U,V;Y_2)+\delta_n(\epsilon),\label{eq:constraint3}
\end{align}
where $V(t):=Y^{t-1}_2$, $t=1,2,\ldots,n$, and $V:=(V(T),T)$.
Furthermore, in light of \eqref{eq:ss},
\begin{align}
	\frac{1}{n}\mathbb{E}[d^{(n)}_{\mathrm{sym}}(S^n,\hat{S}^{(n)})]&\geq\frac{1}{n}\sum\limits_{t=1}^nH(S(t)|Y^n_2)\nonumber\\
	&\geq\frac{1}{n}\sum\limits_{t=1}^nH(S(t)|U(t),V(t),Y_2(t))\nonumber\\
	&=H(S|U,V,Y_2).\label{eq:constraint4}
\end{align}
 It is easy to verify that the joint distribution of $p_{UVXY_1Y_2S}$ factorizes as $p_{UVX}p_{Y_1Y_2S|X}$. In view of \eqref{eq:constraint1}, \eqref{eq:Y_1}, \eqref{eq:constraint3}, and \eqref{eq:constraint4},
\begin{align}
	\overline{C}_{\mathrm{sym}}(D)\leq&\min\{I(X;Y_1),I(X;Y_1|U)+I(U;Y_2),\nonumber\\
	&\hspace{-0.1in}I(X;Y_1,Y_2|U,V)+I(U,V;Y_2)\}+\delta_n(\epsilon)+\epsilon
\end{align}
and
\begin{align}
	D\geq H(S|U,V,Y_2)-\epsilon.
\end{align}
Sending $n\rightarrow\infty$, $\epsilon\rightarrow 0$, and invoking the compactness of the space of $p_{UVXSY_1Y_2}$  proves \eqref{eq:ssupper}.

Next consider the case of sequence-wise logarithmic loss (i.e., $d^{(n)}=d^{(n)}_{\mathrm{seq}}$). Here, we set $V(t):=(Y^{t-1}_2,S^{t-1})$, $t=1,2,\ldots,n$, instead, while all other auxiliary random variables are still identified in the same manner. 
Substituting $Y_1(t)$ with $(Y_1(t),Y_2(t),S(t))$, $t=1,2,\ldots,n$, in the derivation of \eqref{eq:Y_1} yields 
\begin{align}
	\frac{1}{n}\log|\mathcal{M}|\leq I(X;Y_1,Y_2,S|U,V)+I(U,V;Y_2)++\delta_n(\epsilon).\label{eq:replace1}
\end{align}
Moreover, in light of \eqref{eq:ws},
\begin{align}
	\frac{1}{n}\mathbb{E}[d^{(n)}_{\mathrm{seq}}(S^n,\hat{S}^{(n)})]&\geq\frac{1}{n}H(S^n|Y^n_2)\nonumber\\
	&\geq\frac{1}{n}\sum\limits_{t=1}^nH(S(t)|U(t),V(t),Y_2(t))\nonumber\\
	&=H(S|U,V,Y_2).\label{eq:replace2}
\end{align}
Now one can readily establish \eqref{eq:wsupper}
by following the same steps as in the proof of \eqref{eq:ssupper}, with \eqref{eq:replace1} and \eqref{eq:replace2} replacing
\eqref{eq:constraint3} and \eqref{eq:constraint4}.

\section{Proof of Theorem \ref{thm:binary}}\label{app:binary}

The claim $C_{\mathrm{B}}(D)=\overline{R}_{\mathrm{B}}(D)$ is a direct consequence of the fact $C_{\mathrm{B}}(D)\geq\overline{R}_{\mathrm{B}}(D)$ and the following extremal inequality: For $\lambda\geq 0$ and $p_{UXY_1Y_2S}$ factorized as $p_{UX}p_{Y_1Y_2S|X}$, 
\begin{align}
&I(X;Y_1|U)+I(U;Y_2)-\lambda H(S|U,Y_2)\nonumber\\
&\leq 1-H_b(\beta_1)-\lambda H_b(\beta_S)+\max\limits_{\alpha\in[0,\frac{1}{2}]}(H_b(\alpha*\beta_1)\nonumber\\
&\quad-(1-\lambda)H_b(\alpha*\beta_2*\beta_S)-\lambda H_b(\alpha*\beta_2)),\label{eq:upperbound}
\end{align}
which can be proved by observing that
\begin{align}
&I(X;Y_1|U)+I(U;Y_2)-\lambda H(S|U,Y_2)\nonumber\\
&=H(Y_1|U)-H(Z_1)+H(Y_2)-H(Y_2|U)\nonumber\\
&\quad-\lambda (H(S|U)+H(Y_2|U,S)-H(Y_2|U))\nonumber\\
&=H(Y_1|U)-H(Z_1)+H(Y_2)-H(Y_2|U)\nonumber\\
&\quad-\lambda (H(S)+H(X+Z_2|U)-H(Y_2|U))\nonumber\\
&\leq 1-H_b(\beta_1)-\lambda H_b(\beta_S)+H(Y_1|U)-(1-\lambda)H(Y_2|U)\nonumber\\
&\quad-\lambda H(X+Z_2|U)\nonumber\\
&=1-H_b(\beta_1)-\lambda H_b(\beta_S)+\sum\limits_{u\in\mathcal{U}}p_U(u)(H(Y_1|U=u)\nonumber\\
&\quad-(1-\lambda)H(Y_2|U=u)-\lambda H(X+Z_2|U=u))\nonumber\\
&\leq 1-H_b(\beta_1)-\lambda H_b(\beta_S)+\max\limits_{u\in\mathcal{U}}(H(Y_1|U=u)\nonumber\\
&\quad-(1-\lambda)H(Y_2|U=u)-\lambda H(X+Z_2|U=u))\nonumber\\
&\leq 1-H_b(\beta_1)-\lambda H_b(\beta_S)+\max\limits_{p_X}(H(Y_1)-(1-\lambda)H(Y_2)\nonumber\\
&\quad-\lambda H(X+Z_2))\nonumber\\
&=1-H_b(\beta_1)-\lambda H_b(\beta_S)+\max\limits_{\alpha\in[0,\frac{1}{2}]}(H_b(\alpha*\beta_1)\nonumber\\
&\quad-(1-\lambda)H_b(\alpha*\beta_2*\beta_S)-\lambda H_b(\alpha*\beta_2)).
\end{align}

Let $J(\alpha):=H_b(\alpha*\beta_1)-(1-\lambda)H_b(\alpha*\beta_2*\beta_S)-\lambda H_b(\alpha*\beta_2)$. We shall show that when $\beta_1\in(0,\beta_2)$, the function $J(\alpha)$ has a unique maximizer over $[0,\frac{1}{2}]$, which implies $R_{\mathrm{B}}(D)$ is concave in $D$ by \cite[Lemma 5]{Yu21}. 
A direct calculation yields
\begin{align}
	\frac{\partial J(\alpha)}{\partial\alpha}&=(1-2\beta_1)\log\left(\frac{1-\alpha*\beta_1}{\alpha*\beta_1}\right)\nonumber\\
	&-(1-\lambda)(1-2(\beta_2*\beta_S))\log\left(\frac{1-\alpha*\beta_2*\beta_S}{\alpha*\beta_2*\beta_S}\right)\nonumber\\
	&\quad-\lambda(1-2\beta_2)\log\left(\frac{1-\alpha*\beta_2}{\alpha*\beta_2}\right),\\
	\frac{\partial^2 J(\alpha)}{\partial\alpha^2}&=\frac{\phi(\alpha)\log e}{\psi(\alpha)},
\end{align}
where $\phi(\alpha):=\mu\alpha^2-\mu\alpha+\nu$ with
\begin{align}
	\mu&:=\lambda(\beta_2-\beta_2*\beta_S)(1-\beta_2-\beta_2*\beta_S)(1-2\beta_1)^2\nonumber\\
	&\quad-(\beta_1-\beta_2*\beta_S)(1-\beta_1-\beta_2*\beta_S)(1-2\beta_2)^2,\\
	\nu&:=-\lambda(1-\beta_1)\beta_1(\beta_2-\beta_2*\beta_S)(1-\beta_2-\beta_2*\beta_S)\nonumber\\
	&\quad+(1-\beta_2)\beta_2(\beta_1-\beta_2*\beta_S)(1-\beta_1-\beta_2*\beta_S),\label{eq:nu}
\end{align}
and $\psi(\alpha):=(1-\alpha*\beta_1)(\alpha*\beta_1)(1-\alpha*\beta_2*\beta_S)(\alpha*\beta_2*\beta_S)(1-\alpha*\beta_2)(\alpha*\beta_2)$.
Since 
$\left.\frac{\partial J(\alpha)}{\partial\alpha}\right|_{\alpha=\frac{1}{2}}=0$, it suffices to prove that $\frac{\partial J(\alpha)}{\partial\alpha}$  either first decreases and then increases or remains monotonic over $[0,\frac{1}{2}]$, which is equivalent to  $\frac{\partial^2 J(\alpha)}{\partial\alpha^2}$ being initially negative and then positive or maintaining the same sign over $[0,\frac{1}{2}]$. Clearly,  $\psi(\alpha)>0$ for $\alpha\in[0,\frac{1}{2}]$, while $\phi(\alpha)$ is a quadratic polynomial with symmetric about $\alpha=\frac{1}{2}$. If $\mu\leq 0$, then $\frac{\partial^2 J(\alpha)}{\partial\alpha^2}$ obviously possesses the desired property. If $\mu>0$, we have
\begin{align}
	\lambda<\frac{(\beta_1-\beta_2*\beta_S)(1-\beta_1-\beta_2*\beta_S)(1-2\beta_2)^2}{(\beta_2-\beta_2*\beta_S)(1-\beta_2-\beta_2*\beta_S)(1-2\beta_1)^2}.\label{eq:lambda}
\end{align}
Substituting \eqref{eq:lambda} into \eqref{eq:nu} gives
\begin{align}
	\nu&<\frac{(\beta_2-\beta_1)(1-\beta_1-\beta_2)(\beta_1-\beta_2*\beta_S)(1-\beta_1-\beta_2*\beta_S)}{(1-2\beta_1)^2}\nonumber\\
	&<0.
\end{align}
Therefore, $\frac{\partial^2 J(\alpha)}{\partial\alpha^2}<0$ for $\alpha\in[0,\frac{1}{2}]$, confirming that it again satisifies the desired property.


By symmetry, when $\beta_1\in(\beta_2,\beta_2*\beta_S)$, the function $J(\alpha)$ has a unique minimizer over $[0,\frac{1}{2}]$, which in turn implies that $R_{\mathrm{B}}(D)$ is convex in $D$. 
Finally, we observe that $R_{\mathrm{B}}(D)$ is a linear function of $D$ when $\beta_1=\beta_2$.

\section{Proof of Theorem \ref{thm:Gaussian1}}\label{app:Gaussian}

It suffices to show that in case 3), 
\begin{align}
	I(X;Y_1|U)+I(U;Y_2)\leq R_{\mathrm{G}}(D,P)\label{eq:tight}
\end{align}
for $p_{UXY_1Y_2S}$ that factorizes as $p_{UX}p_{Y_1Y_2S|X}$ and satisfies
\begin{align}
	h(S|U,Y_2)\leq D.\label{eq:hD}
\end{align}

Without loss of generality, we can write $Z_2=Z_1+\tilde{Z}$, where $\tilde{Z}\sim\mathcal{N}(0,N_2-N_1)$ is independent of $Z_1$. 
Note that
\begin{align}
	&I(X;Y_1|U)+I(U;Y_2)\nonumber\\
	&=h(Y_1|U)-h(Y_1|X)+h(Y_2)-h(Y_2|U)\nonumber\\
	&=h(Y_1|U,Y_2)-h(Y_1|X)+h(Y_2)-h(Y_2|Y_1)\nonumber\\
	&= h(Y_1|U,Y_2)-h(Z_1)+h(Y_2)-h(\tilde{Z}+S)\nonumber\\
	&\leq h(Y_1|U,Y_2)-\frac{1}{2}\log\left(\frac{2\pi eN_1(N_S+N_2-N_1)}{P+N_2+N_S}\right)\label{eq:sub}
\end{align}
and
\begin{align}
	h(S|U,Y_2)=h(S-Y_2|U,Y_2)=h(X+Z_2|U,Y_2),
\end{align}
which, together with \eqref{eq:hD}, implies
\begin{align}
	h(X+Z_2|U,Y_2)\leq D.\label{eq:invoke}
\end{align}
Now we need the following extremal inequality:
\begin{align}
	h(Y_1|U,Y_2)&\leq \frac{1}{2}\log\left(\frac{(N_2+N_S-N_1)^2}{N^2_S}2^{2h(X+Z_2|U,Y_2)}\right.\nonumber\\
	&\hspace{0.2in}\left.-\frac{2\pi e(N_2-N_1)(N_2+N_S-N_1)}{N_S}\right).\label{eq:extremalGaussian}
\end{align}
Substituting \eqref{eq:extremalGaussian} into \eqref{eq:sub} and invoking \eqref{eq:invoke} proves \eqref{eq:tight}.

It remains to prove \eqref{eq:extremalGaussian}. Let 
\begin{align}
	\Theta_1&:=Z_1-\mathbb{E}[Z_1|Z_2+S]\nonumber\\
	&=Z_1-\frac{N_1}{N_2+N_S}(Z_2+S),\\
	\Theta_2&:=Z_2-\mathbb{E}[Z_2|Z_2+S]\nonumber\\
	&=Z_2-\frac{N_2}{N_2+N_S}(Z_2+S).
\end{align}
It can be verified that $\Theta_1\sim\mathcal{N}(0,\frac{N_1(N_2+N_S-N_1)}{N_2+N_S})$ and $\Theta_2\sim\mathcal{N}(0,\frac{N_2N_S}{N_2+N_S})$ are independent of $(U,X,Y_2)$. We have
\begin{align}
	&h(Y_1|U,Y_2)\nonumber\\
	&=h\left(\left.Y_1-\frac{N_1}{N_2+N_S}Y_2\right|U,Y_2\right)\nonumber\\
	&=h\left(\left.\frac{N_2+N_S-N_1}{N_2+N_S}X+\Theta_1\right|U,Y_2\right)\nonumber\\
	&=h(X+\tilde{\Theta}_1|U,Y_2)+\log\left(\frac{N_2+N_S-N_1}{N_2+N_S}\right)\label{eq:comb1}
\end{align}
and
\begin{align}
	&h(X+Z_2|U,Y_2)\nonumber\\
	&=h\left(\left.X+Z_2-\frac{N_2}{N_2+N_S}Y_2\right|U,Y_2\right)\nonumber\\
	&=h\left(\left.\frac{N_S}{N_2+N_S}X+\Theta_2\right|U,Y_2\right)\nonumber\\
	&=h(X+\tilde{\Theta}_2|U,Y_2)+\log\left(\frac{N_S}{N_2+N_S}\right),\label{eq:comb2}
\end{align}
where $\tilde{\Theta}_1:=\frac{N_2+N_S}{N_2+N_S-N_1}\Theta_1\sim\mathcal{N}(0,\frac{N_1(N_2+N_S)}{N_2+N_S-N_1})$ and $\tilde{\Theta}_2:=\frac{N_2+N_S}{N_S}\Theta_2\sim\mathcal{N}(0,\frac{N_2(N_2+N_S)}{N_S})$. Since $h(X+\tilde{\Theta}_1|U,Y_2)$ and $h(X+\tilde{\Theta}_2|U,Y_2)$ depend on $(\tilde{\Theta}_1,\tilde{\Theta}_2)$ only through their marginal distributions, we can write $\tilde{\Theta}_2=\tilde{\Theta}_1+\Theta'$, where $\Theta'\sim\mathcal{N}(0,\frac{(N_2-N_1)(N_2+N_S)^2}{(N_2+N_S-N_1)N_S})$ is independent of $\tilde{\Theta}_1$. It follows by the entropy power inequality that
\begin{align}
&h(X+\tilde{\Theta}_2|U,Y_2)\nonumber\\
&\geq\frac{1}{2}\log\left(2^{2h(X+\tilde{\Theta}_1|U,Y_2)}+2^{2h(\Theta')}\right)\nonumber\\
&=\frac{1}{2}\log\left(2^{2h(X+\tilde{\Theta}_1|U,Y_2)}+\frac{2\pi e(N_2-N_1)(N_2+N_S)^2}{(N_2+N_S-N_1)N_S}\right).\label{eq:comb3}
\end{align}
Combining \eqref{eq:comb1}--\eqref{eq:comb3} proves \eqref{eq:extremalGaussian}.

\end{document}